# Optimizing Prognostic Biomarker Discovery in Pancreatic Cancer Through Hybrid Ensemble Feature Selection and Multi-Omics Data

**Authors**: John Zobolas[1,2], Anne-Marie George[3], Alberto López[1,2], Sebastian Fischer[4,5], Marc Becker[4], Tero Aittokallio[1,2]

**Affiliations**: [1] Department of Cancer Genetics, Institute for Cancer Research, Oslo University Hospital, Oslo, Norway [2] Oslo Centre for Biostatistics and Epidemiology (OCBE), Department of Biostatistics, University of Oslo, Oslo, Norway [3] Department of Informatics, University of Oslo, Oslo, Norway [4] Department of Statistics, Ludwig Maximilian University of Munich, Munich, Germany, [5] Munich Center for Machine Learning (MCML)

## Abstract

Prediction of patient survival using high-dimensional multi-omics data requires systematic feature selection methods that ensure predictive performance, sparsity, and reliability for prognostic biomarker discovery. We developed a **hybrid ensemble feature selection** (hEFS) approach that combines data subsampling with multiple prognostic models, integrating both embedded and wrapper-based strategies for survival prediction. Omics features are ranked using a voting-theory-inspired aggregation mechanism across models and subsamples, while the optimal number of features is selected via a Pareto front, balancing predictive accuracy and model sparsity without any user-defined thresholds. When applied to multi-omics datasets from three pancreatic cancer cohorts, hEFS identifies significantly fewer and more stable biomarkers compared to the conventional, late-fusion CoxLasso models, while maintaining comparable discrimination performance. Implemented within the open-source **mlr3fselect** R package, hEFS offers a robust, interpretable, and clinically valuable tool for prognostic modelling and biomarker discovery in high-dimensional survival settings.

# Background

High-dimensional omics data are increasingly generated in patient cohorts, presenting both opportunities and challenges for biomarker discovery [McDermott2013, Rufeng2022]. A major challenge is the typical **"p >> n"** setting, where the number of molecular features (p) vastly exceeds the number of available patient samples (n). In cancer research, this curse of dimensionality problem is further complicated by right-censored survival outcomes, making statistical analysis and model validation particularly difficult [Turkson2021, Riley2024]. Additional challenges arise when integrating multi-omics data with machine learning (ML), where ensuring the prognostic model interpretability, reliability, and generalizability remains a critical concern [He2010, Olivier2019]. For example, in pancreatic cancer clinical management, identifying robust prognostic biomarkers is critical for patient stratification, treatment planning and outcome prediction [Halbrook2023, Tripathi2024, Passaro2024]. However, limited sample sizes, heterogeneous disease progression, and scarce high-quality datasets make predictive modeling and biomarker identification particularly challenging. Addressing these challenges requires robust feature selection (FS) and predictive modeling approaches tailored to survival analysis to support clinical applications.

For biomarker discovery to be clinically applicable, three key criteria must be met: (1) ***predictivity*** - the selected molecular features must yield high predictive performance on independent patient data, ensuring that the prognostic model generalizes beyond the training cohort and avoids overfitting, (2) ***sparsity*** - a minimal set of non-redundant, informative features must be selected to make the clinical implementation of the prognostic model more feasible, and (3) ***stability*** - the selected features must be robust to data perturbations and consistently associated with clinical outcomes across data splits and sample cohorts, to avoid false discoveries that occur due to technical noise inherent to high-throughput omics profiling [Kalousis2006, Hedou2024, Theng2024]. In high-dimensional settings, additional FS objectives include reducing model training time, identifying clinically cost-effective predictors, and improving interpretability by linking features to biological mechanisms [Guyon2003, Saeys2007].

Feature selection methods can be broadly categorized into **embedded, filter-based, and wrapper-based** approaches, each with unique strengths and limitations [Theng2024]. Embedded methods, such as **CoxLasso** [Tibshirani1997], incorporate FS directly into the model training, but they can be sensitive to hyperparameter tuning and sample variability [Fan2001, Meinshausen2010]. Filter-based methods, like mRMR [Peng2005] and variance filtering, rank features independently of a predictive model, offering scalability, but they often require *ad hoc* filtering thresholds [Saeys2007, Asir2016, Seijo-Pardo2019]. Wrapper-based methods, such as random survival forest-based recursive feature elimination (**RSF-RFE**) [Ishwaran2008, Pang2012], iteratively evaluate feature subsets based on their predictive performance, hence improving model interpretability at the cost of higher computational complexity. While FS for classification and regression tasks has been extensively studied [Guyon2002, Fan2002, Zhang2007, Hofner2014, Saqib2019, Theng2024], adaptations to survival outcomes remain scarce, due to the additional complexity introduced by censoring and time-to-event data.

To mitigate FS instability in high-dimensional settings [Kalousis2006], **homogeneous** ensemble approaches based on data subsampling or other types of perturbations have been developed, such as Stability Selection [Meinshausen2010] and its adaptations to survival models [Laimighofer2016, Yin2017, Kahn2019, Liang2023], which offer a probabilistic control over false discoveries. **Heterogeneous** ensemble feature selection methods, which aggregate results from diverse FS algorithms [Sarkar2021, Asghar2024], have shown to reduce method-specific biases and improve selection robustness [Zhang2019, Pes2020]. More recently, **hybrid ensemble FS** (hEFS) strategies that leverage data perturbation and method diversity have emerged in classification tasks, identifying more stable and biologically meaningful biomarker signatures [Colombelli2022, Budhraja2023, Claude2024]. However, dedicated hEFS frameworks optimized for survival analysis — balancing sparsity, predictive accuracy, and stability without relying on user-defined thresholds — are still lacking. This gap underscores the need for automated and robust FS methods, tailored specifically to high-dimensional prognostic biomarker discovery.

The aforementioned challenges become even more pronounced in **multi-omics survival analysis,** due to increased data heterogeneity, variable omics dimensionalities, and potential missingness patterns [Zhao2024]. Standard integration strategies, such as **early fusion** and **late fusion,** have been adapted to multi-omics settings [Ding2022]. Early fusion combines all omic layers into a joint optimization framework, capturing potential cross-omic interactions, but it often neglects differences in omics data distributions. Late fusion maintains the individual aspects of each omic layer by training separate models and combining their predictions post-hoc, which reduces the risk of overfitting, but it lacks explicit modeling of cross-omic relationships. More sophisticated approaches, including regularization techniques like **priority-Lasso** [Klau2018], **IPF-Lasso** [Boulesteix2017], and **sparse Bayesian hierarchical models** [Zhao2024], attempt to bridge this gap. However, recent benchmarking studies have shown that predictive performance often declines as more omic layers are added [Wissel2023, Li2024], reflecting the persistent "**p >> n**" problem and the need for survival FS strategies that are noise-resistant and ensure stability, sparsity, and interpretability during multi-omics integration.

To address these challenges, we developed a hEFS method designed for robust and interpretable biomarker discovery in high-dimensional and multi-omics survival settings. Our approach combines data subsampling with nine survival prediction models, combining both embedded and wrapper-based FS strategies. Feature rankings are calculated using a voting-theory-inspired mechanism, treating each model-subsample pair as a voter and the molecular features as candidates [Drotar2019, Lackner2023]. To determine the optimal number of features, we introduce a Pareto front-based strategy [Das1999], which balances **predictive performance and model sparsity**, eliminating the need for any user-defined thresholds. We extend the hEFS framework to multi-omics data via a late-fusion strategy, and benchmark it against conventional late-fusion CoxLasso on three pancreatic cancer patient cohorts. Our results demonstrate that the proposed hEFS method achieves **comparable discrimination performance** while selecting **significantly fewer and more stable features**. The methodology is implemented in the open-source `mlr3fselect` **R-package** [Becker2025], built upon the modular mlr3 ecosystem [Lang2019], which provides a comprehensive toolset for machine learning in R, including support for data resampling, model tuning, and performance evaluation, along with a seamless use of diverse models across survival, classification, and regression tasks.

## Results

A Pareto-Driven Framework for Hybrid Ensemble Feature Selection (hEFS)

To address the challenge of identifying sparse, stable, and generalizable biomarker signatures from high-dimensional omics data, we developed a Pareto-driven hybrid ensemble feature selection (hEFS) framework (**Fig. 1**). hEFS integrates data perturbation, model diversity, voting-based feature ranking, and automated Pareto-based selection of the feature subset size that best balances predictive performance and model sparsity. This enables a systematic discovery of stable, interpretable, and parsimonious prognostic biomarkers in high-dimensional settings. The key components of hEFS are as follows:

- **Data and model diversity**: hEFS combines data perturbation - through repeated random subsampling of the patient cohort - with a library of heterogeneous predictive models, generating all possible data-model combinations used for feature selection and performance evaluation (**Fig. 1b**).
- **Flexible feature selection via embedded or wrapper methods**: To derive sparse feature subsets from diverse models and data resamplings, the models that implement embedded FS select features during their model fitting (e.g., penalized regression), while other models (e.g. random forest) are wrapped in a

Recursive Feature Elimination (RFE) framework (**Fig. 1c**). In RFE, an inner cross-validation loop evaluates model performance at each iteration, and the feature subset corresponding to the best-performing iteration is selected as the output. For each data-model pair, feature selection is performed on the training set, while predictive performance is evaluated on a separate hold-out test set, providing an unbiased performance estimate (**Fig. 1d**).

- **Robust feature ranking via voting**: hEFS aggregates feature selections and performance scores across all data-model pairs using a voting-theory-inspired Satisfaction Approval Voting (SAV) mechanism, where data-model pairs act as voters and the selected features represent their chosen candidates [Drotar2019, Lackner2023]. SAV normalizes approval scores, i.e., number of models for which the feature was selected, by the total number of features selected per model. We use a weighted version of SAV that incorporates the performance scores from the test sets to weight the models' influence in the vote (**Methods**). Thus, we are favoring features identified by sparser, higher-performing models. This results in a robust and transparent consensus ranking that reflects both selection frequency and model quality (**Fig. 1e**).
- **Automated selection of the final feature set**: A Pareto front of model sparsity (number of selected features) versus predictive performance is computed across all data-model pairs. A knee-point identification (KPI) method [Das1999] locates the "knee" on this Pareto front, representing the best trade-off between sparsity and accuracy. The corresponding number of features at this knee point—derived automatically and without any user-defined thresholds—is then used to filter the voting-based ranking, resulting in a minimal, high-performing biomarker panel for downstream prognostic modeling (**Fig. 1e**).

Additional technical optimizations designed to enhance biomarker discovery include:

1) **One-SE rule optimization:** To favor sparsity, RFE selects the smallest feature subset with predictive performance within one standard error (SE) from the best-performing iteration [Hastie2009, Kuhn2013, Chen2021]. This approach systematically prioritizes sparser models during the RFE process (**Fig. 1c**).
2) **Beta-distribution-driven RFE subset sizing:** To favor the evaluation of smaller feature subsets, hEFS samples subset sizes from a skewed Beta distribution, reflecting the assumption that most features in high-dimensional omics data are noisy. This concentrates the search in the lower-dimensional feature space, where stable and sparse signatures are more likely to be found.
3) **Flexible integration:** The framework accommodates classification, regression, and right-censored survival outcomes; a wide range of models supporting embedded or importance-based FS (e.g., random forests, lasso, SVMs, boosting); diverse resampling strategies (e.g., out-of-bag error for random forests); and various performance measures (e.g., accuracy, AUC, C-index).

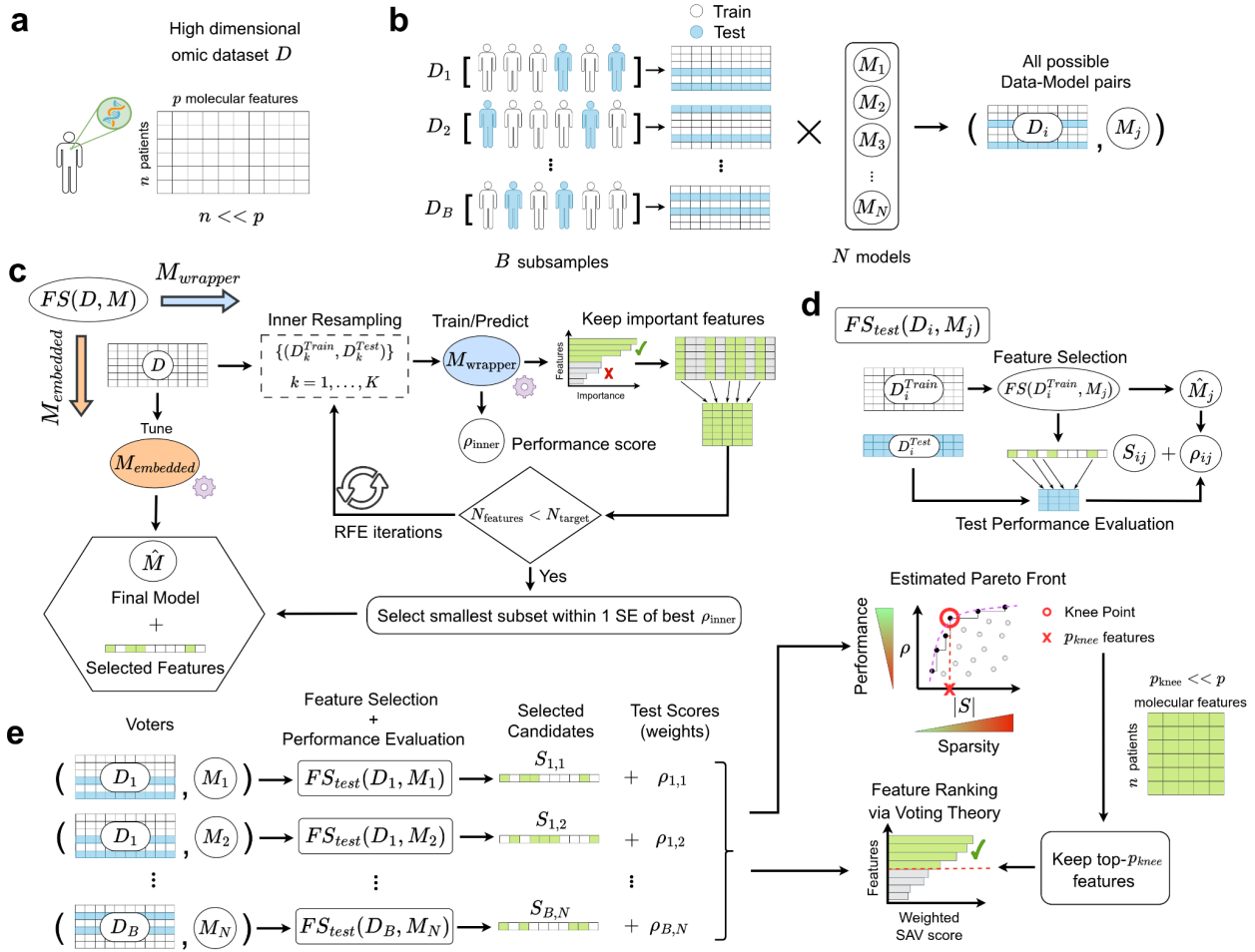

**Figure 1: Overview of the hybrid ensemble feature selection (hEFS) framework**. (a) High-dimensional omics dataset measured across a cohort of n patients, where the goal is to identify a subset of prognostic molecular biomarkers among the p features. (b) Data perturbation and model diversity: the dataset is subsampled $B$ times, and each subsample is paired with one of the $N$ different models, forming $B \cdot N$ unique data-model combinations. (c) Feature selection (FS): models with embedded FS select features as part of their model fitting and hyperparameter tuning, while models that produce per-feature importance scores are used in a wrapper-based recursive feature elimination (RFE) framework, which iteratively removes features and selects the subset whose performance lies within one standard error of the best cross-validated performance ($\rho_{inner}$). (d) For each data-model combination, feature selection is performed on the training set via step (c), resulting in a trained model and a selected feature subset $S_{ij}$. These features are then evaluated on the hold-out test set to obtain unbiased performance estimates $\rho_{ij}$. (e) hEFS aggregates all selected feature subsets $S_{ij}$ and performance scores $\rho_{ij}$ across data-model pairs using a voting-theory-inspired Satisfaction Approval Voting (SAV) mechanism, which ranks features by assigning higher scores to those selected by high-performing, sparser models. A

Pareto front is computed over the model sparsity (number of features $S_{ij}$) and predictive performance ($\rho_{ij}$), identifying a "knee point" that optimally balances these two objectives. The final hEFS output consists of the top-ranked features at this knee point, yielding a sparse, high-performing, and interpretable prognostic biomarker signature.

This generalized design makes hEFS applicable across a broad range of omics datasets and outcome prediction tasks (e.g. binary response class prediction or time-to-event survival analysis), regardless of the data distributions, and ensures robustness for complex multi-omics biomarker discovery tasks.

The full pipeline is implemented in R within the **mlr3** ecosystem [Lang2019]. It is fully customizable, and easily scalable as each data-model pair runs independently in parallel based on available cores. A tutorial for the wrapper-based hEFS variant is also available online [Zobolas2025]. For further implementation details, see **Methods**.

## Application of hEFS to multi-omics PDAC datasets

### A Comprehensive Framework for Evaluation of Multi-Omics Feature Selection Strategies

We applied hEFS to three retrospective multi-omics cohorts of patients diagnosed with pancreatic ductal adenocarcinoma (PDAC) and right-censored survival outcomes, spanning different disease stages and omic modalities [Raphael2017, Cao2021, Osipov2024]. Two of the datasets included only early-stage (stage I–II) PDAC patients, while the third [Cao2021] encompassed all stages (I–IV), enabling evaluation across clinically diverse populations. PDAC is one of the most aggressive solid tumors, with limited treatment options and poor prognosis, underscoring the urgent need for robust prognostic biomarkers to guide clinical decision-making and stratify patients for personalized therapies [Loosen2017, Khomiak2020]. All three PDAC cohorts represent **high-dimensional "p >> n" settings**, with between 70 and 125 patients and thousands of molecular features per sample, hence providing challenging real-world prediction scenarios. The datasets differ in the number and types of omic modalities, as well as the number, scale, and distribution of features per data type - ranging from continuous gene expression profiles to sparse, discrete mutation count data. Median survival times ranged between 20 and 30 months, while differences in censoring patterns provided complementary evaluation settings for benchmarking feature selection strategies (**Sup. Fig. 1**). Key dataset characteristics are summarized in **Table 1**. Clinical features common to all cohorts include age, sex, and tumor stage. Additional preprocessing details, including patient inclusion criteria, feature filtering, and the rationale for modality selection, are described in the **Methods** section.

**Table 1**. Overview of the three multi-omics PDAC cohorts used in this study.

| Study | CPTAC-PDAC [Cao2021] | TCGA-PDAC [Raphael2017, Wissel2023] | MolTwin [Osipov2024] |
|---|---|---|---|
| **Number of patients (events)** | 125 (70) | 81 (48) | 71 (47) |
| **Censoring rate*** | 44% | 41% | 34% |
| **Omic data types (number)** | GEX, CNVs, Proteomics, Phosphoproteomics, N-glycoproteomics (5) | GEX, CNVs, Mutations, Methylation, RPPA (5) | SNVs, CNVs, Indels, Digital Pathology (4) |
| **Number of clinical features** | 7 | 4 | 10 |
| **Total number of features[+]** | 10,007 | 8,194 | 1,319 |

*The MolTwin cohort has patients administratively censored at 72 months, while for the TCGA and CPTAC datasets the censoring is distributed across the study time. [+]Includes all molecular and clinical features.

We extended hEFS to multi-omics analysis using a **two-stage late fusion strategy** [El-Manzalawy2018] (**Fig. 2**). In the first stage, feature selection is performed independently for each omic layer using only the training cohort of a given multi-omics dataset. This results in a set of omic-specific biomarkers selected via the hEFS procedure, which combines SAV-based feature ranking with Pareto front-based selection of the final feature subset size (**Fig. 1e**). In the second stage, the selected features from all the omic layers are concatenated into a joint feature matrix to form a unified multi-omics biomarker signature, which is used to train a predictive model and evaluate its performance on the test cohort. This approach avoids cross-omic information leakage during feature selection, preserves omic-level interpretability of the selected biomarkers, and has been shown to leverage the most predictive modalities more effectively than early fusion approaches [Wissel2023]. Beyond the current study, this modular late-fusion framework also provides a flexible and scalable template for benchmarking feature selection strategies in multi-omics settings. By decoupling omics-specific feature selection from model training and allowing independent, resampling-based evaluation, it accommodates heterogeneous data types, varying response outcomes (e.g. classification or survival), various evaluation metrics, and multiple integration approaches—making it broadly applicable to future benchmarking efforts.

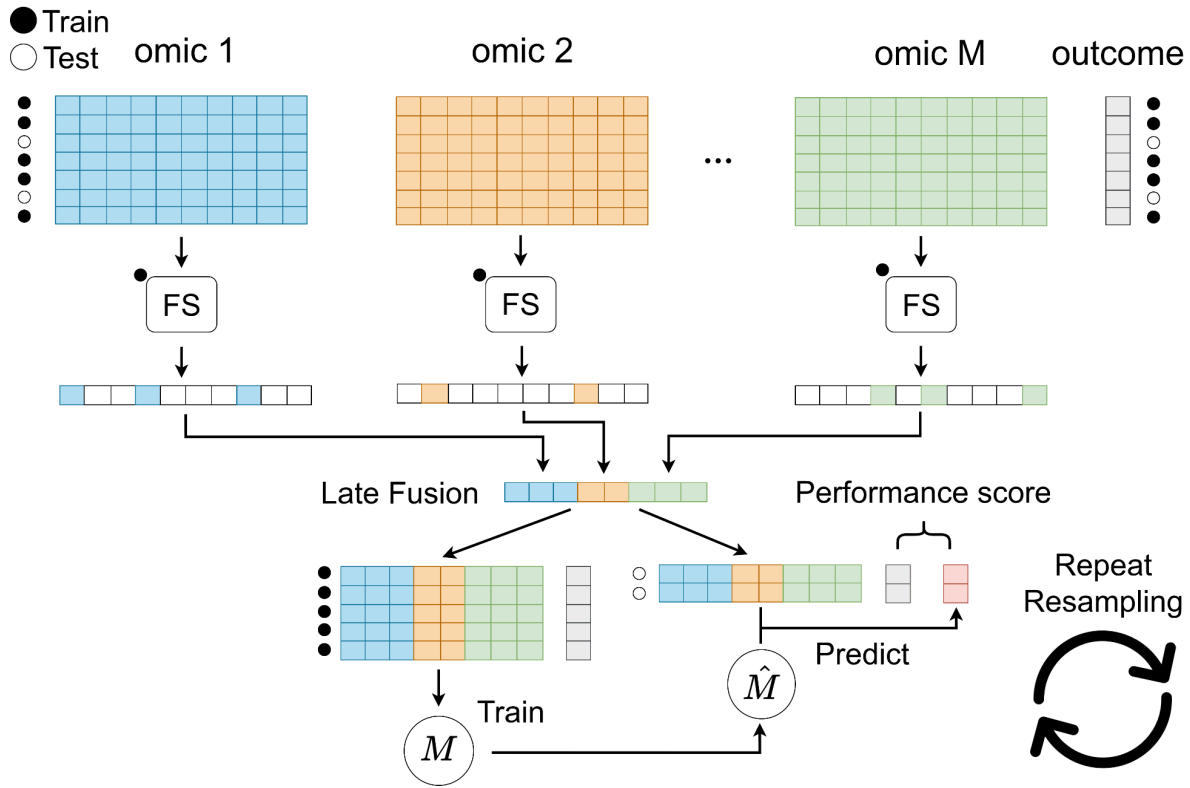

**Figure 2: Late-fusion framework for multi-omics feature selection and model evaluation.** Each omic modality undergoes separate feature selection (FS) using only the training data. The resulting omic-specific feature subsets are then concatenated into a joint feature matrix - representing the multi-omics biomarker signature - which is used to train a predictive model ($M$) and evaluate its performance on the corresponding test data. This procedure is repeated across multiple resampling iterations to assess selection stability, sparsity and predictive accuracy. Clinical and surgical pathology variables (e.g., age, sex, tumor stage) are always included in the final model without feature selection, as they are few in number and represent known prognostic factors. To provide a baseline reference, the framework includes evaluation of models trained solely on clinical variables, enabling the assessment of the added prognostic value from multi-omics integration.

To ensure robust evaluation, we performed 100 Monte Carlo cross-validation (MC-CV) iterations for each multi-omics PDAC dataset, using an 80/20 stratified train/test split. Stratification was based on censoring status to preserve the proportion of censoring observations across resampling runs. For the CPTAC-PDAC dataset [Cao2021], we additionally stratified by tumor stage, as all four stages were available. MC-CV provides a fixed test set size and helps mitigate bias in downstream assessments of feature selection stability and redundancy. Within each split, feature selection was applied separately to each omic layer using one of the following four methods (corresponding to the FS step in **Fig. 2,** where $N$ denotes the number of distinct models used, and $B$ the number of subsamples per model as illustrated in **Fig. 1a**):

1. **CoxLasso:** A baseline approach using a single regularized Cox model with embedded feature selection, applied without subsampling ($N = 1$, $B = 0$) [Tibshirani1997];
2. **hEFS (9 models):** A full ensemble using nine diverse survival models ($N = 9$), combined with subsampling ($B = 100$);
3. **hEFS (3 RSFs):** A reduced ensemble using only three random survival forest variants ($N = 3$, $B = 100$) [Ishwaran2008, Jaeger2023];
4. **EFS (CoxLasso):** A homogeneous ensemble using CoxLasso applied to multiple subsamples ($N = 1$, $B = 100$), serving as a non-hybrid reference method.

CoxLasso provides a straightforward baseline, while the other three approaches represent different hEFS variants offering different trade-offs between ensemble diversity and computational cost. The RSF-based ensemble leverages out-of-bag error estimates during RFE optimization (**Fig. 1c**), reducing runtime. The homogeneous EFS-CoxLasso variant is computationally lighter and generally more stable than applying CoxLasso directly without prior subsampling.

For multi-omics model integration (denoted as model $M$ in **Fig. 2**), we employed **BlockForest** [Hornung2019], a variant of random survival forests (RSF) [Ishwaran2008] that accounts for group structure in multi-omics data. This choice was motivated by prior findings showing that group-aware integration methods outperform group-naive approaches in multi-omics settings [Herrmann2021, Wissel2023]. Clinical pathology variables (e.g., age, sex, tumor stage) were always included in the final model without prior feature selection, as they represent established prognostic markers, and were treated equivalently to omics-derived features during the model integration (i.e., no forced prioritization or weighting was applied). For reference, we also evaluated RSF models trained solely on clinical features, which consistently outperformed Cox proportional hazards models across all datasets (**Sup. Fig. 2**). While it is well recognized that the choice of integration model and modality composition can affect downstream predictive performance [Wissel2023, Li2024], our primary objective was to systematically compare **feature selection strategies**, rather than optimize the integration architecture itself.

We benchmarked each method across the following five key dimensions, which reflect essential criteria for general-purpose feature selection evaluation:

- **Sparsity**: Number of selected features per omic layer.
- **Stability**: Similarity of selected feature subsets across resampling iterations, assessed using the Nogueira similarity metric [Nogueira2018], which adjusts for random chance agreement and differences in subset sizes.
- **Redundancy**: Proportion of significantly redundant feature pairs (FDR-adjusted $p < 0.05$) within each omic layer, based on the rank correlation coefficient ξ [Chatterjee2021]; redundancy rate was also quantified using mean absolute correlation.
- **Predictivity**: Assessed using Harrell's C-index [Harrell1982], which measures the discriminatory ability of the selected biomarker signature to differentiate high- versus low-risk patients. The C-index was computed on models trained using the late-fused clinical + multi-omics feature matrix (**Fig. 2**), as well as on the clinical-only RSF reference model.
- **Computational cost:** Quantified as the runtime of each FS method per training set.

A detailed overview of the benchmark design, including model tuning, inner resampling (performed entirely within each training set to avoid data leakage) and the complete list of hEFS models, is provided in the **Methods** section.

hEFS Improves Sparsity in Biomarker Selection Across PDAC Cohorts

We first evaluated the sparsity of the biomarker signatures selected by each feature selection method, both at the level of individual omics and the overall fused multi-omics panel. **Fig. 3a** displays the number of selected features per omic modality across the three PDAC cohorts. The CoxLasso baseline consistently yielded the highest number of selected features, with **median counts often exceeding 100**, particularly for copy number variation (CNV) data in the CPTAC and TCGA cohorts. Other modalities with relatively high number of biomarkers included mutations in the TCGA cohort and N-glycoproteomics and proteomics in the CPTAC cohort, each with median selections around 50 features. This likely reflects the tendency of Lasso to retain many weakly predictive features when signal strength is low and distributed across many variables, especially in modalities such as CNV and mutations, which often consist of discrete or sparse data [Zou2005]. In contrast, the hEFS variants achieved markedly sparser selections across all omics and cohorts. Throughout the 100 Monte Carlo cross-validation iterations, both hEFS (9 models) and hEFS (3 RSFs) selected **less than 15 features per omic**, regardless of the initial feature dimensionality (see

**Methods**). This trend is more clearly visible in **Sup. Fig. 3**, where the CoxLasso baseline is omitted to better visualize the hEFS feature count distributions.

**Fig. 3b** summarizes the total number of selected features across all omic layers and clinical variables for each cohort. While overall sparsity varied by cohort, **hEFS (3 RSFs) was the most parsimonious approach** in most cases, followed by EFS (CoxLasso) and hEFS (9 models). For example, in the CPTAC and TCGA datasets, CoxLasso selected 200–300 multi-omics biomarkers out of >8,000 total candidates, whereas **hEFS methods typically retained ~50**. An exception was the hEFS (9 models) configuration in the TCGA cohort, which selected a higher number of mutation features (**Fig. 3a**). This was driven by Pareto front configurations that included model–data pairs with larger feature sets, effectively shifting the knee point—the basis for biomarker selection—toward solutions with more features (**Sup. Fig. 4**). In the MolTwin cohort—where the dimensionality was lower (1,319 features)—CoxLasso selected ~100 multi-omics biomarkers, while hEFS variants again reduced this to ~50 or fewer. Notably, CoxLasso exhibited **higher variance** in feature counts across resampling iterations and PDAC cohorts, compared to hEFS, suggesting **less consistent selection behavior** (**Fig. 3b**). This highlights a potential stability advantage of the ensemble-based hEFS strategy, which more reliably favors sparser biomarker signatures through its Pareto-front selection procedure (**Fig. 1e**).

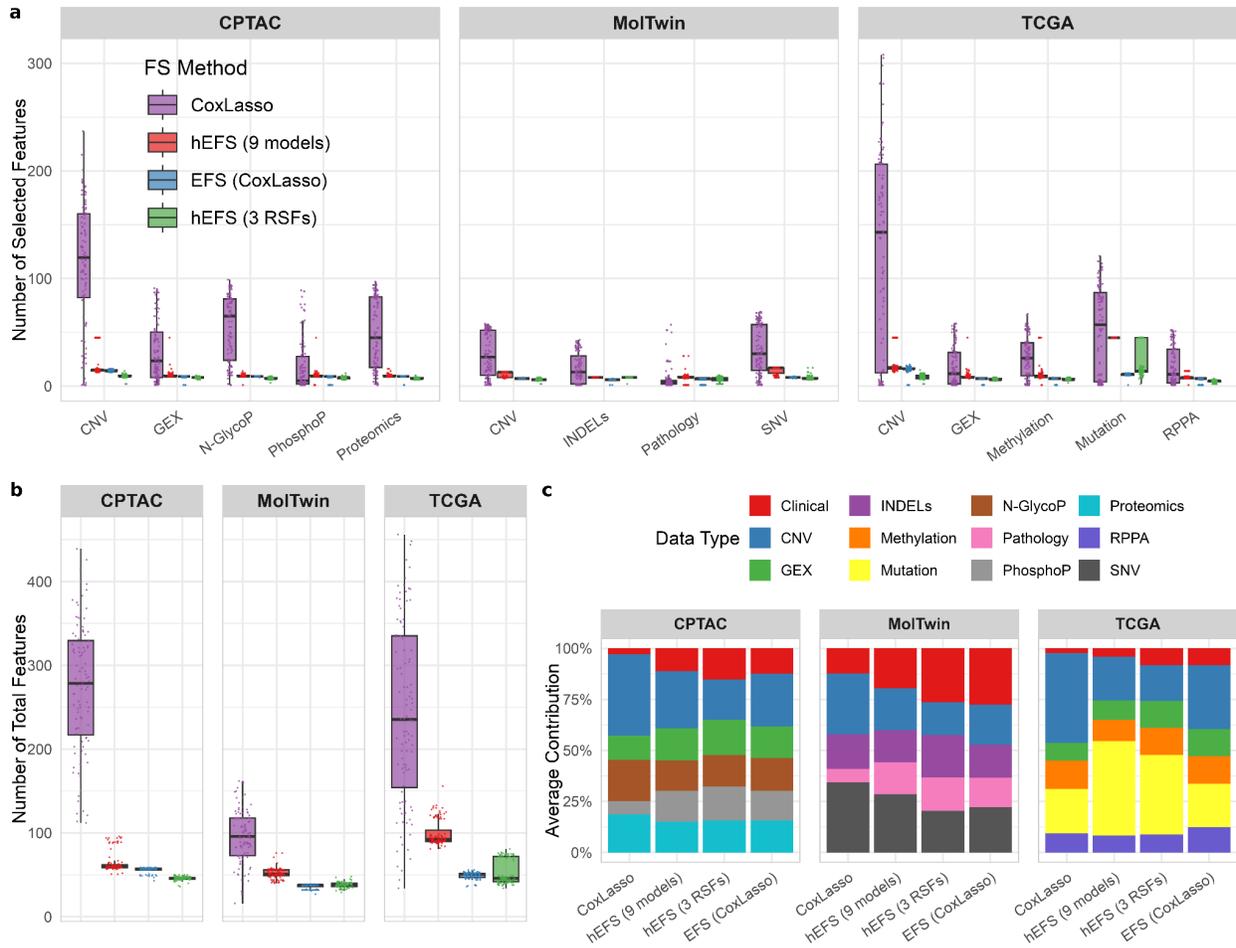

**Figure 3: Sparsity and modality composition of selected biomarker signatures across feature selection methods and PDAC cohorts.** (a) Number of selected features per omic modality for each dataset and feature selection method, evaluated over 100 Monte Carlo cross-validation (MC-CV) iterations. hEFS variants consistently yield sparser selections compared to the CoxLasso baseline. (b) Total number of features selected across all modalities (including clinical variables) across the 100 MC-CV iterations. hEFS (3 RSFs) is the most parsimonious method overall, with CoxLasso showing high variance and larger feature counts. (c) Relative contribution of each omic type to the final multi-omics biomarker signature. While all methods preserve modality diversity, ensemble-based approaches show more balanced inclusion patterns. Color coding corresponds to data types shown in the legend.

Finally, **Fig. 3c** shows the composition of the multi-omics biomarker signatures by data type. Since each FS method yielded different total numbers of selected features (**Fig. 3b**), the stacked bar plots represent relative contributions, not absolute feature counts. Clinical variables were always included and thus appear across all methods. Despite these differences, we observe that all omic modalities are represented in the final signatures across methods and datasets, demonstrating that the **late-fusion design preserves modality diversity** (**Fig. 2**). Notably, the hEFS (3 RSFs) and EFS (CoxLasso) variants produced more balanced modality contributions, avoiding the overrepresentation of CNV or mutation data seen with CoxLasso and hEFS (9 models) methods in the TCGA and CPTAC cohorts. This pattern is influenced by both data-type characteristics and the geometry of the estimated Pareto front, where outlier models with larger selected feature sets on some inner resamplings — more likely with more models — push the knee point toward higher feature numbers, increasing the final biomarker set size (**Sup. Fig. 4**).

hEFS Enhances Selection Stability Without Compromising Redundancy

We next evaluated the stability and redundancy of the selected feature subsets across the 100 MC-CV iterations (**Fig. 4**). Stability was assessed using the Nogueira similarity [Nogueira2018], an unbiased metric that corrects for random agreement between feature subsets with different sizes. All hEFS variants exhibited **consistently higher similarity scores** than the baseline CoxLasso across omic types and datasets (**Fig. 4a**), supporting prior findings that ensemble methods improve biomarker selection reproducibility [Pes2020]. Notably, even the homogeneous ensemble of CoxLasso models – EFS (CoxLasso) – benefited from subsampling, achieving improved stability over the single-model baseline. The lower stability of CoxLasso is partly due to its tendency to select larger subsets (**Fig. 3**), but more critically, it reflects the challenge of detecting weak signals in high-dimensional, noisy omics data—leading to inconsistent feature selection across the MC-CV folds.

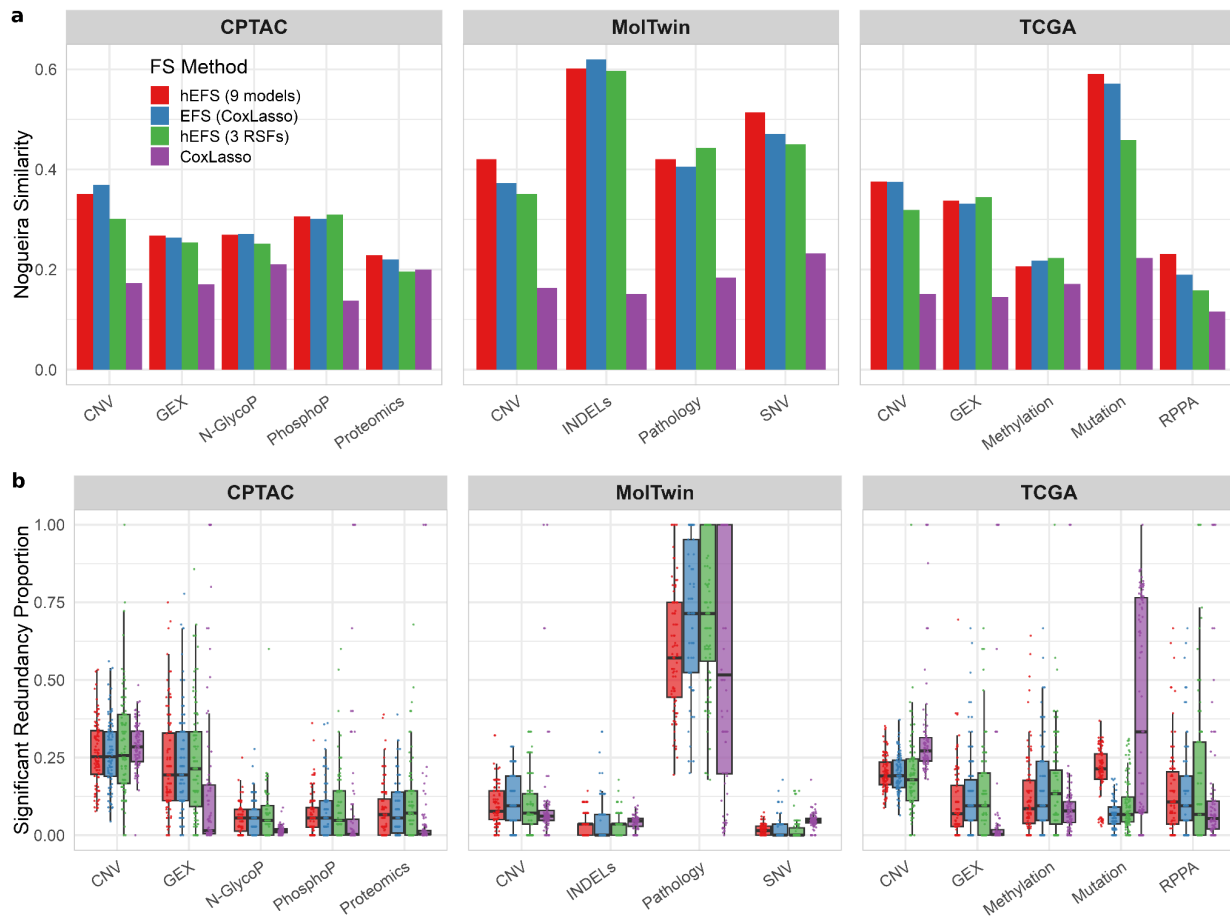

**Figure 4. Stability and redundancy of selected biomarker signatures across feature selection methods and PDAC cohorts.** (a) Nogueira similarity [Nogueira2018] quantifying the stability of selected features across 100 Monte Carlo cross-validation (MC-CV) iterations, stratified by omic type and dataset. Higher values indicate greater similarity, corrected for subset size differences. Ensemble-based approaches consistently achieved higher stability than the CoxLasso baseline. (b) Proportion of significantly redundant feature pairs within each omic type, based on Chatterjee's ξ correlation [Chatterjee2021] with FDR-adjusted $p < 0.05$. Lower values reflect less redundancy among selected features. Boxplots summarize variability across the MC-CV iterations.

Stability varied across omic types and cohorts. Among hEFS variants, hEFS (9 models) was typically the most stable in certain omic types (e.g., CNV in MolTwin, proteomics in CPTAC, RPPA in TCGA), although practical differences relative to hEFS (3 RSFs) and EFS (CoxLasso) were generally small. The MolTwin cohort exhibited the highest overall stability, likely driven by its lower initial dimensionality (most omics had < 300 pre-filtered features, e.g., only 63 INDELs; see **Methods**) and feature distributions that were predominantly discrete with many zeros. A similar effect was seen in the TCGA mutation data, where stability was relatively high compared to the continuous omic types from the same dataset—here, the small number of non-zero entries in the mutation data constrained the set of features that can be consistently selected across MC-CV iterations. Overall, absolute stability values remained modest (typically <0.4), in line with prior observations that feature selection stability is inherently low in high-dimensional settings [Dernoncourt2014]. Finally, Nogueira stability scores proved robust; for each FS method–omic–dataset combination, we randomly sampled 50 MC-CV runs (and their corresponding selected feature subsets) from the original 100 runs, and repeated this resampling 50 times. Interquartile ranges of the resulting stability scores were consistently below 0.05 (**Sup. Fig. 5**). These results indicate that the observed stability patterns are not driven by specific MC-CV splits and would likely persist even with fewer iterations, reflecting consistent and reproducible differences in stability across FS methods and omic types.

In terms of redundancy, the three hEFS variants produced highly similar results across omics and PDAC cohorts, while CoxLasso exhibited slightly greater variation. Across all FS methods, fewer than 30% of selected feature pairs showed statistically significant dependence (FDR-adjusted $p < 0.05$) in most omic types and datasets (**Fig. 4b**). Redundancy was quantified using the ξ correlation [Chatterjee2021], a flexible measure of dependence capable of capturing bi-directional, linear and nonlinear associations between continuous and categorical variables - making it well-suited for heterogeneous multi-omics data. Compared to Pearson or Spearman correlation (**Sup. Fig. 6**), ξ correlation generally identified less significant dependencies, reflecting its sensitivity to more general forms of statistical dependence, rather than strict linear or monotonic associations—and it also exhibited lower variability across omic types and datasets. Interestingly, CoxLasso exhibited in some cases lower significant redundancy proportion values (**Fig. 4b**), particularly in gene expression and proteomics data. This can be attributed to its lower stability (**Fig. 4a**); since selected feature subsets vary substantially across the MC-CV folds, the pooled feature set becomes more diverse across the feature space. As a result, fewer feature pairs co-occur frequently enough to be detected as significantly redundant. This reflects an *artifact of instability,* rather than a genuine feature decorrelation, and underscores the importance of interpreting redundancy metrics alongside stability patterns.

Mean absolute correlation scores across the MC-CV iterations remained low (median ≤ 0.2; **Sup. Fig. 7**), indicating low-to-moderate redundancy rates across the feature selection methods, regardless of data type. A notable exception was the **pathology** modality in the MolTwin cohort, which exhibited higher redundancy rates across all methods. This was expected given the inherent collinearity among handcrafted nuclear descriptors - statistical summaries (e.g., percentiles, standard deviations) derived from AI-extracted morphology features - as previously described in Osipov et al. [Osipov2024].

## hEFS Balances between Predictivity and Computational Efficiency

Discriminatory performance was evaluated using Harrell's C-index, by comparing feature selection (FS) strategies on clinical plus late-fused multi-omics feature matrices against a clinical-only RSF baseline. Across all PDAC datasets, we found that **the choice of FS method had a minimal impact on the discriminatory performance** (**Fig. 5a**). Specifically, C-index values ranged between **0.54–0.56** in the TCGA cohort, **0.60–0.62** in the MolTwin cohort, and **0.58–0.60** in the CPTAC cohort. The relatively higher performances seen in the CPTAC and MolTwin cohorts are likely attributable, respectively, to their larger sample size (**Table 1**) and careful preprocessing of genomic and pathology features to emphasize biologically relevant signals [see **Methods**; Osipov2024]. Importantly, **none of the FS approaches improved discriminatory performance beyond the clinical-only reference model**, in line with prior observations [Herrmann2021, Wissel2023].

We next asked to what extent predictive outcomes were shaped by the choice of integration model. When using group-naïve RSF, the performance after hEFS was broadly comparable to BlockForest (**Sup. Fig. 8a**), suggesting that once sparsity and stability are introduced through FS, even simpler integration strategies perform relatively well, given adequate sample sizes. In contrast, CoxLasso as an integration model generally underperformed regardless of the FS strategy, deteriorating relative to the baseline in the CPTAC cohort and performing worse than random in the TCGA cohort (**Sup. Fig. 8b**). Only in the MolTwin cohort, where features had already undergone extensive preprocessing and were limited due to prior FS (<100; **Fig. 1b**), CoxLasso achieved moderate performance. This likely reflects the sensitivity of CoxLasso to collinearity and high-dimensional noise, which is only partially mitigated by FS.

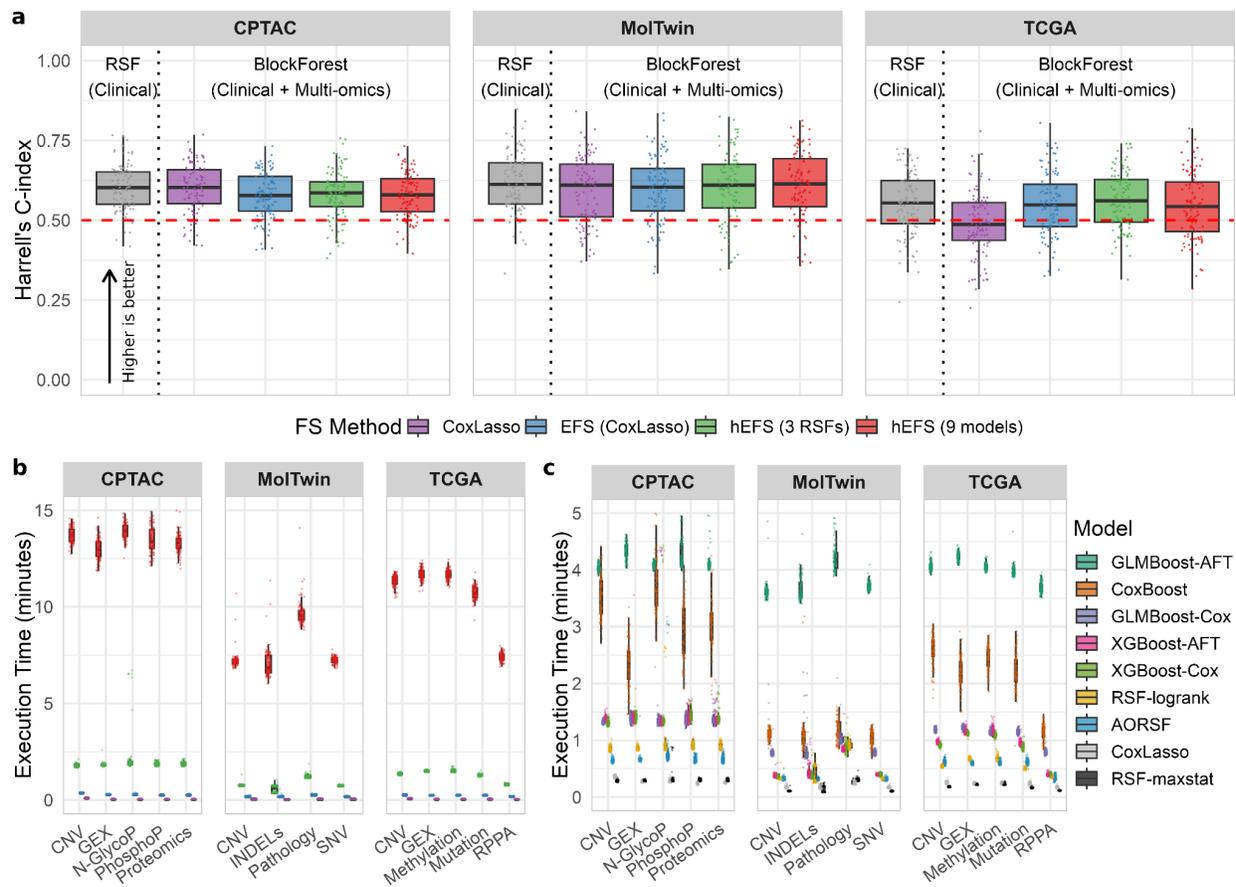

**Figure 5. Predictive performance and computational cost across feature selection methods and PDAC cohorts.** (a) Predictive performance of BlockForest models that integrate clinical data with late-fused, feature-selected multi-omics matrices, benchmarked against a baseline RSF trained only on clinical features (grey). Harrell's C-index shows that feature selection strategy had a little influence on predictivity, and none of the methods outperformed the clinical-only baseline. The red dashed line indicates random discriminatory ability (C-index = 0.5). (b) Execution times for each FS strategy, stratified by cohort and omic type. Runtimes reflect cohort size and feature dimensionality. (c) Breakdown of full hybrid ensemble (hEFS, 9 models) runtimes by individual method. Each method is a homogeneous ensemble with B = 100 subsamples. GLMBoost-AFT and CoxBoost were the most computationally demanding, while CoxLasso and RSF-based approaches were the fastest. All results were obtained using parallelization on 40 cores on the Fox HPC cluster (**Methods**).

Importantly, FS remained critical even when using a group-aware method like BlockForest; completely omitting FS in the TCGA cohort caused performance to drop to near-random levels (C-index ≈ 0.5), when combining all omics with clinical data, whereas applying hEFS restored it back to baseline (**Sup. Fig. 9**). In the CPTAC cohort, in contrast, the impact of FS was less pronounced, possibly because BlockForest captures feature interactions across omics types that are partially lost when applying sparse hEFS per omic; this highlights that using FS prior to integration is not always a clear-cut decision and may involve a trade-off between sparsity and interaction modeling. Interestingly, using CoxLasso for feature selection in combination with BlockForest integration resulted in random discriminatory ability (**Fig. 5a**), underscoring that mismatches between FS strategies and integration models can severely impair predictive performance. However, introducing improved stability through subsampling corrected this issue in the ensemble CoxLasso variant (EFS), which restored the performance to baseline levels—highlighting stability as a decisive factor for achieving reliable outcomes in both FS and omics integration.

In contrast, **the choice of predictive modalities had a stronger impact on performance** than the integration model itself. Focusing on clinical data combined with gene expression (GEX)—a combination previously shown to be effective [Zhao2015, Hornung2019, Vale-Silva2021, Wissel2023]—substantially improved discriminatory performance. In both the CPTAC and TCGA cohorts, integrating merely GEX (where hEFS with all 9 models was applied) with clinical features, rather than all the available omics, increased the **C-index to ~0.64** (**Sup. Fig. 9**). Interestingly, even using GEX alone with RSF (without integration with clinical data) exceeded the predictive value of the clinical-only RSF baseline, likely reflecting RSF's ability to capture non-linear effects and interactions within the transcriptomic data. This indicates that GEX captures significant prognostic information, consistent with recent information-theoretic analyses in TCGA data identifying molecular profiles—including GEX, CNV, and mutation data—as primary modalities carrying task-relevant survival signals [LiangP2023]. Taken together, these results indicate that prognostic performance in PDAC is shaped both by modality choice and by the FS/integration strategy, but relative gains remain modest. The primary role of FS in this context is not to boost the predictive accuracy, but to reduce dimensionality, promote sparsity, and enhance interpretability in clinical applications, while retaining clinically relevant predictive signals.

In terms of computational cost, execution times varied considerably across the FS methods and cohorts (**Fig. 5b–c**, **Sup. Table 1**). The full hybrid ensemble FS (hEFS with 9 models) was by far the most resource-intensive, requiring up to ~14 minutes per training set on the CPTAC cohort, ~8 min in MolTwin (fewer features per omic), and ~11 min in the TCGA cohort. All runs were parallelized on 40 cores with peak memory usage reaching ~40 GB, using the Fox HPC cluster (4 interactive nodes, each with 2× AMD EPYC 7702 CPUs and 1 TB RAM; **Methods**). These differences largely reflected the cohort sample sizes (**Table 1**). Stratification into individual models (**Fig. 5c**) revealed that GLMBoost-AFT was the slowest (~4 minutes per training set), followed by CoxBoost (~2–4 minutes depending on dataset size). GLMBoost-Cox and the two XGBoost variants (Cox, AFT) required ~1 minute each, while AORSF and RSF-logrank took ~30–40 seconds. The fastest were CoxLasso and RSF-maxstat, each completing in ~13–15 seconds. Notably, each of these models within the hEFS framework is itself a homogeneous ensemble trained with $B$ = 100 subsamples; thus, the reported execution times already reflect tuning and fitting; for example, 100 CoxBoost models per training set. Within datasets, runtime was strongly influenced by omic dimensionality; for example, in the TCGA cohort, all omics except for RPPA had 2000 features and required ~11 minutes, whereas RPPA (190 features) completed in ~7.5 minutes. Similarly, the Pathology modality in the MolTwin cohort, which had the largest feature set (794 features), consistently required more execution time than the other omics.

We note that users have the flexibility in selecting which base models to include in the hybrid ensemble. For instance, removing the heaviest component, GLMBoost-AFT —which accounts for roughly one-third of total hEFS (9 models) runtime—can substantially reduce the computational burden without markedly affecting overall performance, highlighting that ensemble size and composition can be adjusted to optimize the trade-off between efficiency and predictive accuracy. Among reduced ensembles (**Sup. Table 1**), hEFS (3 RSFs) consistently required only 1–2 minutes, benefitting from the efficient use of out-of-bag error during RFE iterations, instead of costly inner cross-validation. EFS (CoxLasso) required ~15 seconds, making it far faster than the hybrid hEFS variant, while standalone CoxLasso with no subsampling was nearly instantaneous (<3 seconds on average). Overall, execution time scaled with the number of patients, number of features, and ensemble size, but "smart" hEFS configurations, such as the RFE-based feature selection using random survival forests, provided substantial time savings without loss of predictive performance.

## Discussion

Pancreatic ductal adenocarcinoma (PDAC) remains one of the most lethal malignancies, with poor survival rates and limited therapeutic options [Siegel2025]. Accurate prognostic biomarkers are urgently needed to guide patient stratification and treatment decisions as early as possible [Loosen2017, Pishvaian2020, Khomiak2020, Passaro2024]. Advances in high-throughput molecular and genetic profiling have enabled the generation of multi-omics data with unprecedented resolution, but the high dimensionalities, small sample sizes, and heterogeneity of PDAC cohorts make biomarker discovery particularly challenging in censored survival settings. These challenges highlight the need for fully-automated and stable feature selection (FS) strategies tailored to survival analysis, which can balance predictive performance with sparsity and interpretability in integrative multi-omics models.

A wide range of FS strategies have been proposed for high-dimensional survival analysis, spanning sparsity-promoting embedded approaches such as **CoxLasso** [Tibshirani1997] and related regularization or boosting techniques [Fan2002,

Zhang2007, Binder2008, Simon2011, Hofner2014], computationally efficient filter methods [Welchowski2019, Bommert2022], and wrapper-based approaches like recursive elimination with random survival forests [Ishwaran2008, Pang2012]. Homogeneous ensemble methods that perturb data to stabilize FS—such as **SurvRank** [Laimighofer2016], stability selection for Cox-Lasso [Yin2017], or **VSOLassoBag** [Liang2023]—have demonstrated improved sparsity and more reliable FS compared to single-model approaches. Furthermore, several single-omic methods have been extended to the multi-omics survival domain, including **priority-Lasso** [Klau2018], **IPF-Lasso** [Boulesteix2017], and **sparse Bayesian hierarchical models** [Zhao2024], although most of these approaches rely on *ad hoc* thresholds or strong structural assumptions, such as prior knowledge of the order of modality blocks. Thus, despite progress, survival-specific ensemble FS frameworks tailored for multi-omics integration remain underdeveloped.

We addressed this gap with a novel **hybrid ensemble feature selection (hEFS)** framework that integrates data perturbation, heterogeneous base models, and voting-theory-inspired feature ranking with automated Pareto-based knee-point identification for selecting the number of features. Unlike many existing methods, hEFS requires **no user-specified thresholds or predefined feature counts**, instead determining the optimal subset size directly from the data. This design ensures a balance between sparsity and predictivity, accommodates censored outcomes across diverse omics, and improves stability through subsampling and model diversity. Thus, hEFS functions as a **fully automated and reproducible FS framework** that enables interpretable and clinically relevant multi-omics biomarker discovery in high-dimensional survival settings.

To evaluate hEFS and other multi-omics FS approaches in a clinically meaningful context, we implemented a two-stage late-fusion benchmark that systematically assessed FS across multiple dimensions. As Frank E Harrell noted, "*a molecular signature can be either parsimonious or predictive, but not both*" [Harrell2022], yet real-world biomarker discovery requires balancing additional aspects, including stability, redundancy, and computational cost. Our late-fusion design first performs FS independently within each omic modality, before integrating the selected features into a unified multi-omics signature, thereby avoiding cross-omic information leakage and preserving modality-level interpretability. Coupled with repeated resampling, this design enabled reproducible quantification of all key FS criteria, hence providing a general framework and systematic blueprint for future benchmarking studies that emphasize clinically relevant, interpretable, and efficient biomarker panels.

Applying this framework to three PDAC multi-omics cohorts—the largest benchmark of its kind to date—demonstrated the practical benefits of hEFS for survival-focused feature selection. Across diverse omic modalities and censoring patterns, **hEFS consistently identified sparser and more stable biomarker panels** than baseline CoxLasso, selecting on average ~10 features per omic versus ~60 using CoxLasso, with substantially lower variance. Importantly, sparsity and stability gains did not compromise predictive performance, which matched a clinical RSF baseline. While absolute discrimination in PDAC remained modest (C-index rarely exceeding 0.6), redundancy among selected features was generally low, and subsampling substantially improved selection stability. The ensemble size could be adjusted to match available computational resources without notable loss in performance, and the choice of integration model (e.g., BlockForest vs RSF) had less impact once stability was enforced. Instead, **modality choice was the main driver of performance**, with gene expression combined with clinical variables providing the best discriminatory value (C-index ~0.64), in line with prior evidence [Zhao2015, Hornung2019, Vale-Silva2021, Wissel2023]. Overall, these findings underscore that hEFS's principal value lies not in maximizing raw predictive accuracy— that is still modest in PDAC—but in producing **reproducible, interpretable, and clinically actionable multi-omics biomarker signatures** suitable for high-dimensional survival analysis.

Our study has some limitations that should be acknowledged. First, we did not include deep learning (DL) models or nonlinear dimensionality reduction methods (e.g., PCA, Autoencoders). While such approaches can boost predictive performance in multi-omics settings [Poirion2021, Chen2023, Wissel2023], they typically transform the feature space, thereby sacrificing direct interpretability of the selected biomarkers. Moreover, the high computational demands of DL methods, especially for model tuning, make them less suited for integration into an ensemble FS framework. Our focus was therefore on methods that preserve original features and provide interpretable multi-omics signatures. Similarly, although histopathology images have shown synergistic value when integrated with molecular profiles on some TCGA datasets [LiangP2023], our scope was limited to openly accessible tabular molecular data or preprocessed imaging features. Second, while our benchmark represents the largest PDAC-specific multi-omics ML study to date, we were unable to validate biomarker signatures on an independent external cohort. This limitation reflects the scarcity of PDAC datasets with overlapping omic modalities, harmonized preprocessing, and sufficient sample sizes for machine learning analysis. Efforts toward data sharing and harmonization will be critical to fully assess the generalizability of multi-omics biomarker signatures across cohorts.

Looking ahead, several directions for future research emerge. Methodologically, the multi-criteria benchmark framework could be extended to include additional evaluation criteria beyond sparsity, stability, redundancy, predictivity, and computational cost.

**Biological interpretability** remains a key factor, with knowledge-driven discovery strategies leveraging pathways [Thomas2019, Thomas2022], GO annotations [TheGeneOntologyConsortium2023], and molecular interaction networks (e.g., gene–gene or protein–protein) showing promise [He2010, McDermott2013, Wang2022]. **Reliability**—quantifying the proportion of truly informative features—represents another important criterion, which could be systematically investigated using synthetic survival datasets that vary in censoring mechanisms, proportions of informative versus noise features, different data distributions and correlation structures [Giordano2022, Hedou2024]. Hybrid ensemble methods could also be extended to explicitly integrate reliability [Wu2007, Kursa2010, Thomas2017, Ren2023], stability [Bommert2017], redundancy [Capraz2024], or biological priors into the Pareto front optimization, potentially using higher-dimensional fronts and automated knee-point detection techniques [Chiu2016, Guerreiro2020, Li2020]. Another promising direction is to apply subsampling and ensembling strategies within integrative multi-omics models, such as IPF-Lasso [Boulesteix2017, Castel2025] or priority-Lasso [Klau2018], thereby extending the hEFS framework beyond single-modality FS.

A critical step toward clinical translation will be to move from prognostic to predictive biomarker discovery [Passaro2024]. This will require prospective validation in large clinical studies, ideally using datasets that include treatment information from patients with PDAC [Arango-Argoty2025]. By enabling reproducible, interpretable, and clinically relevant biomarker panels, future extensions of hEFS can help bridge the gap between computational biomarker discovery and clinical application. To facilitate adoption, our methodology is openly available in the `mlr3fselect` R-package [Becker2025], ensuring accessibility and ease of use for researchers within the broader mlr3 ecosystem [Lang2019].

# Funding

The project received funding from the European Union's Horizon 2020 research and innovation programme under grant agreement No 101016851, project PANCAIM. TA: Research Council of Finland (grants 340141, 344698, 367855); the Cancer Society of Finland, the Norwegian Cancer Society (grants 216104 and 273810), Norwegian Health Authority South-East (grants 2020026 and 2023105), and the Sigrid Jusélius Foundation.

# Conflict of interest

Authors state no conflict of interest.

## Methods

### Hybrid Ensemble Feature Selection (hEFS) Framework

**Notation**

Let $D \in \mathfrak{R}^{n \times p}$ denote the omics data matrix with $n$ patients and $p$ molecular features, where typically $p \gg n$. The outcome is $Y = (y_1, \ldots, y_n)$. In classification, each $y_i \in \{1, \ldots, C\}$ encodes the class label. In regression, $y_i \in \mathfrak{R}$ represents a continuous response. In single-event right-censored survival analysis, $y_i = (t_i, d_i)$ consists of the observed survival or censoring time $t_i$ and event indicator $d_i \in \{0, 1\}$ (1 = event, 0 = censored). We assume an unknown set of informative features $S^* \subseteq \{1, \ldots, p\}$ with complement the uninformative feature set $\{1, \ldots, p\} \setminus S^*$. The goal of feature selection is to approximate $S^*$.

**Data and model perturbation**

The dataset is randomly subsampled $B$ times, producing training–test splits $(D_i^{train}, D_i^{test})$, $i = 1, \ldots, B$. Each subsample is paired with each of the $N$ predictive models $\{M_j\}_{j=1}^{N}$. This yields $K = B \cdot N$ data–model pairs, indexed by $k = (i, j)$.

**Model-specific feature selection**

For each data-model pair $(D_i, M_j)$, training on $D_i^{train}$ yields a fitted model $\widehat{M_j}$ and a model-specific feature subset $S_{ij}$ (**Fig. 1c-d**):

1. **Embedded selection:** If $M_j$ supports embedded feature selection (e.g., Lasso, Cox-Lasso), training directly produces the set of selected features $S_{ij}$ and the fitted model $\widehat{M_j}$. Models may tune internal regularization parameters (e.g., the penalty $\lambda$ in Lasso) via inner cross-validation.
2. **Wrapper-based recursive feature elimination (RFE):** If $M_j$ provides feature importance scores (e.g., Random Forests, boosting), RFE is applied on $D_i^{train}$. At each iteration, a subset of features is removed, the model is refit, and performance ($\rho_{inner}$) is assessed by inner cross-validation, or out-of-bag error for Random Forests. Iterations continue until the number of features is less than or equal to a predefined target ($N_{target}$). The final subset $S_{ij}$ (and corresponding fitted model $\widehat{M_j}$) is chosen as the smallest feature set within one standard error of the best-performing iteration [Hastie2009, Kuhn2013], favoring sparsity without compromising performance.

After the feature selection step, the output trained model $\widehat{M_j}$ is evaluated on the $D_i^{test}$, restricted to the selected features $S_{ij}$, producing an unbiased performance score $\rho_{ij}$.

To determine the subset sizes explored during RFE, we use a Beta-distribution–based sampling scheme. This generates a stochastic decreasing sequence of candidate subset sizes **biased toward smaller feature sets** in high-dimensional settings. Formally, probabilities are defined by a rescaled $Beta(\alpha = 0.5, \beta)$ distribution over feature counts. The parameter $\beta$ (shape) controls the degree of skewness toward smaller subsets, with practical defaults chosen according to the dimensionality of the dataset (e.g., fewer, more heavily skewed sizes for $p > 10^3$). A separate sampling parameter controls the number of output subsets and therefore the number of RFE iterations. For illustration, in a task with $p = 2000$ features, the generated sequence with 15 RFE iterations and chosen $\beta = 20$ might be {2000, 431, 172, 159, 133, 121, 92, 56, 41, 30, 25, 22, 9, 7, 3}, reflecting the bias toward smaller subsets.

**Voting-based feature ranking**

Each data–model pair $k \in \{1,...,K\}$ (where $K = B \cdot N$, indexing the combination of the $i$-th subsample and the $j$-th model) acts as a voter. Each pair (voter) produces a selected feature set $S_k$ (candidates) with predictive performance $\rho_k$ (weights). For simplicity, we use the single index $k$ to represent all data–model combinations, and we assume that larger weights correspond to higher predictive performance.

We define the **weighted satisfaction approval voting (SAV) score** for feature $i$ as:

$$score_{SAV}(i) = \frac{1}{Z} \sum_{k=1}^{K} \rho_k \cdot \frac{\mathbb{1}\{i \in S_k\}}{|S_k|}$$

where $Z = \sum_{k=1}^{K} \frac{\rho_k}{|S_k|}$ is a normalization factor ensuring that $score(i) \in [0, 1]$, for all features $i$.

This choice of $Z$ provides several advantages:

- The maximum score is exactly 1, allowing direct interpretation of the most frequently and strongly selected features as top-ranking candidates.
- Scores remain proportional to the weighted fraction of voters supporting a feature, while accounting for variable set sizes $|S_k|$ and performance weights $\rho_k$.
- The $[0, 1]$ scale allows the scores to be loosely interpreted as **feature selection probabilities** [Meinshausen2010].

Note that for $|S_k| = 1$, the expression above reduces to the standard **weighted approval voting (AV)**. If all $\rho_k$ weights are equal, then the above formula reduces to the un-weighted score, where each voter contributes equally regardless of predictive performance.

We favor SAV over standard AV in the hEFS framework because it mitigates the "*tyranny of the majority*" where a weak majority of models (e.g., 51%) selects similar features, while all the remaining (49%) models' selection are ignored as they are outweighed by the majority. SAV distributes each model's "approval weight" across its selected features, so sparsely selected but high-performing features receive proportionally higher credit. Like AV, SAV is simple, fast, committee monotone (i.e., allowing more features to be selected without reducing a feature's score), and Pareto-efficient, but additionally it emphasizes stability and avoids over-representing less sparse, low-performance models. These properties—fast computation, committee monotonicity, and Pareto efficiency—make SAV particularly suitable for our setting and preclude the use of other proportional approval voting rules (e.g., Phragmén or Proportional Approval Voting), which prioritize proportional representation of voters rather than stability or feature sparsity [Lackner2023].

**Pareto-based determination of final set size**

Candidate solutions are pairs $(|S_k|, \rho_k)$, representing the trade-off between feature sparsity and performance. The **Pareto front** is estimated by fitting a linear model as $\rho \sim 1/|S|$, predicting performance across feature counts from 1 to $max(|S_k|)$. The **knee point** is defined as the point on the estimated front with maximal perpendicular distance from the line joining its two extremes [Das1999].

The final biomarker panel is obtained by selecting the top $p_{knee}$ features with the highest SAV scores, resulting in the set $S_{hEFS}$. This yields the reduced matrix $D_{final} \in \Re^{n \times p_{knee}}$.

# PDAC Cohorts Preprocessing

**CPTAC**

The CPTAC cohort comprises 140 patients with pancreatic cancer and right-censored survival outcomes [Cao2021]. Raw data files were downloaded from LinkedOmics (http://www.linkedomics.org/data_download/CPTAC-PDAC/). After filtering to retain only PDAC patients (135 patients) and excluding those with missing survival time or tumor stage, **125 stage I-IV patients** remained for analysis.

Clinical variables (7 in total) included age, sex (male: n = 66, female: n = 59), tumor stage, number of lymph nodes examined, number of positive (i.e. metastatic) lymph nodes, and lymphovascular invasion. Tumor stage was collapsed to four categories as IA–IB → 0: n = 23, IIA–IIB → 1: n = 54, III → 2: n = 39 and IV → 3: n = 9. Lymphovascular invasion was collapsed into two categories: 0 for 'not identified' or 'indeterminate' (n = 39) and 1 for 'present' (n = 86). Survival time was converted from days to months, and status was coded as 1 for deceased and 0 for censored. The censoring rate in this cohort was 44%.

Five omic modalities were included: bulk mRNA expression (GEX), copy number variation (CNV), proteomics, phosphoproteomics, and N-glycoproteomics, consistent with **Fig. 7a** in [Cao2021]. Data characteristics were: GEX—positive, normalized counts; CNV—continuous gene-level log2 ratios; proteomics and phosphoproteomics—positive, median-normalized intensities; N-glycoproteomics—TMT log2 ratios (peptide level). Preprocessing steps included retaining tumor samples, matching patient identifiers across modalities, removing features with >10% missing values, and imputing remaining missing values with the median of each feature. To reduce dimensionality, the 2,000 most variable features were retained per modality. After preprocessing, the **six modalities** contained the following number of features: GEX (2,000), CNV (2,000), proteomics (2,000), phosphoproteomics (2,000), N-glycoproteomics (2,000), and clinical variables (7), for a total of **10,007 multi-omics features**.

**MolTwin**

The MolTwin cohort [Osipov2024] includes 74 patients with PDAC and right-censored survival outcomes. Data were obtained from the online publication's data sources ("Source Data Fig. 1"). While the original cohort includes up to 10 modalities, we retained only patients with complete data for at least three modalities and a minimum of 60 patients per selected modality combination, resulting in **71 stage I-II patients** and four omic modalities for analysis: somatic single-nucleotide variants (SNV), copy number variations (CNV), small insertions/deletions (INDELs), image-derived digital pathology features.

Clinical variables (10 in total) included age, sex (male: n = 37, female: n = 34), weight, height, BMI (body mass index), tumor stage (stage I (0): n = 16, stage II (1): n = 55), histological grade (3 classes as 0: n = 6, 1: n = 48, 2: n = 17), lymph invasion (no invasion (0): n = 24, invasion (1): n = 47), clinical site (collapsed from 5 classes as 0: n = 49, >1: n = 22), and histology behavior (2 classes from ICD-O-3 [WHO2000]: '81403' (adenocarcinoma) → 0: n = 12, '85003' (invasive) → 1: n = 59). Survival time was converted from days to months, and status was coded as 1 for deceased and 0 for censored. The censoring rate in this cohort was 34%, with administrative censoring applied at the maximum follow-up time of 72 months.

Genomic features were derived from a targeted 648-gene oncology panel and preprocessed to reduce redundancy by removing constant features and dropping highly correlated features (Spearman correlation > 0.95), following the preprocessing in [Osipov2024]. Data characteristics were as follows: SNVs — categorical values {0,1,…,6}, with most entries being 0; CNVs — continuous gene-level log2 ratios, centered around zero with many zeros; INDELs — categorical values {0,1,2,3}, again predominantly zeros. Pathology features, engineered as statistical summaries (e.g., percentiles, standard deviations) of AI-extracted nuclear morphology and staining texture descriptors as described in Osipov et al., were standardized (z-scored) after removing constant features. After preprocessing, the **five modalities** contained the following number of features: SNVs (274), CNVs (178), INDELs (63), pathology (794), and clinical variables (10), for a total of **1,319 multi-omics features**.

**TCGA**

The TCGA cohort is based on TCGA-PAAD [Raphael2017], as curated by Wissel et al. [Wissel2023] for machine learning–based survival prediction. Raw data were downloaded from Zenodo (https://zenodo.org/records/7529459, file 'preprocessed.zip', containing file 'PAAD_data_preprocessed.csv'). After filtering to retain only PDAC patients (by histological type) and excluding those with missing clinical stage or histological type, we further removed 3 patients with stage III, 1 with stage IV, and 2 with missing stage, leaving **81 stage I–II patients** for analysis. Stages III–IV were excluded due to minority representation and potential bias in survival modeling.

Clinical variables (4 in total) included age, sex (male: n = 46, female: n = 35), tumor stage, and number of metastatic lymph nodes. Tumor stage was collapsed to two categories: IA–IB → 0 (n = 8) and IIA–IIB → 1 (n = 73). Survival time was converted from days to months, and status was coded as 1 for deceased and 0 for censored. The censoring rate in this cohort was 41%.

Five omic modalities were retained: gene expression (GEX), copy number variation (CNV), mutation, DNA methylation, and protein expression (RPPA). The miRNA modality was excluded due to nonsensical data distributions. Data characteristics were: GEX—positive, normalized counts; CNV—discrete values {−2,-1,0,1,2}; RPPA—normalized expression; mutation—non-silent mutation counts (≥0, up to 50, mostly zeros); methylation—beta values in (0,1). Preprocessing steps followed the same strategy as CPTAC, removing features with >10% missing values and imputing the remainder with feature-wise medians. To reduce dimensionality, the 2,000 most variable features were retained per modality (except RPPA, which contained 190 features). After preprocessing, the **six modalities** contained the following number of features: GEX (2,000), CNV (2,000), mutation (2,000), methylation (2,000), RPPA (190), and clinical variables (4), for a total of **8,194 multi-omics features**.

## Multi-omics PDAC Benchmark Design

### Resampling strategy

All PDAC cohorts were evaluated with 100 Monte Carlo cross-validation (CV) splits (80/20 train/test), stratified by censoring status. For CPTAC, stratification also included tumor stage, as this is the only cohort containing stage III–IV patients. Each split defined a training set, on which feature selection was performed within the late-fusion framework (**Fig. 2**), and a held-out test set, on which performance was evaluated. Harrell's concordance index [Harrell1982] was used throughout to compute test-set performance, independent of the integration model.

### Models and Tuning within hEFS

Within each outer Monte Carlo training set, hEFS (or plain CoxLasso) was applied. For all hEFS configurations, we used $B = 100$ subsamples, stratified by censoring status (**Fig. 1b**). Unless otherwise stated, all other model hyperparameters used package defaults.

**Wrapper-based FS Models**

For the random survival forest family (RSF variants, AORSF), we did not tune hyperparameters. The inner performance estimate used during RFE was the out-of-bag error, defined as 1 − Harrell's C-index, so each model was fit once per iteration. For

XGBoost models, inner C-index performance ($\rho_{inner}$) was computed using 5-fold cross-validation.

Feature subset sizes for RFE were generated by a Beta-distribution sampling scheme; the **number of RFE iterations** and **skewness parameter** β were chosen by input dimensionality $p$:

- $p > 1500$: 15 iterations, β = 20
- $500 < p \leq 1500$: 15 iterations, β = 15
- $100 < p \leq 500$: 10 iterations, β = 5
- $10 < p \leq 100$: 10 iterations, β = 3

We implemented this via the *mlr3* callback *clbk("mlr3fselect.rfe_subset_sizes")*. At each RFE run, the one-standard-error rule selected the best iteration, implemented via *clbk("mlr3fselect.one_se_rule")*.

### RSF-logrank

We implemented the Random Survival Forest using the *ranger* R package [Wright2017]. The split criterion was set to `splitrule = "logrank"`. Models were trained with `num.trees = 500` and `importance = "permutation"`.

### RSF-maxstat

We also implemented a Random Survival Forest with *ranger* [Wright2017, WrightM2017] using maximally selected rank statistics as the split criterion. Here we set `splitrule = "maxstat"`, with the same configuration of `num.trees = 500` and `importance = "permutation"` as in RSF-logrank.

### AORSF

The Accelerated Oblique Random Survival Forest was implemented with the *orsf* R package [Jaeger2022]. The model was fit using `control_type = "fast"`, `n_tree = 500` and `importance = "permute"`.

### XGBoost-Cox

We trained gradient-boosted survival trees with the *xgboost* R package [Chen2016], using `objective = "survival:cox"` and `eval_metric = "cox-nloglik"`. Models were run with `nrounds = 500`, `eta = 0.1`, `max_depth = 6`, `booster = "gbtree"`, and `tree_method = "hist"`. Early stopping was applied with

`early_stopping_rounds = 42` using 5-fold inner cross-validation. Internal tuning was implemented with the *mlr3* callback *clbk("mlr3fselect.internal_tuning")*: each CV fold returned an early-stopped number of boosting rounds (the test CV folds acting as validation sets), and the final model was refitted on the full training set with the average of these values.

**XGBoost-AFT**

We also used the Accelerated Failure Time variant of XGBoost [Barnwal2022] using `objective = "survival:aft"` and `eval_metric = "aft-nloglik"`. Training was configured with `nrounds = 500`, `eta = 0.1`, `max_depth = 6`, `booster = "gbtree"`, `tree_method = "hist"`, and `early_stopping_rounds = 42`. The AFT loss was parameterized with `aft_loss_distribution = "logistic"` (log-logistic distribution) and `aft_loss_distribution_scale = 1` ($\sigma = 1$). As with the Cox model, internal tuning via *clbk("mlr3fselect.internal_tuning")* selected the final `nrounds` by averaging early-stopping results across folds.

**Embedded FS Models**

**GLMBoost-Cox**

We implemented the generalized linear survival model using a boosting algorithm via *mboost::glmboost()* with Cox proportional hazards [Hofner2014]. Models were specified with `family = "coxph"` and `center = TRUE`. Hyperparameters `mstop` (boosting rounds) and `nu` (learning rate) were tuned via random search with 25 evaluations using *mlr3*, employing 5-fold inner cross-validation and Harrell's C-index as the performance measure; `mstop` was tuned between 10 and 500, and `nu` between 0 and 0.1.

**GLMBoost-AFT**

The accelerated failure time version of *mboost::glmboost()* [Schmid2008, Hofner2014] was configured with `family = "loglog"` (log-logistic survival distribution) and `center = TRUE`, using the same random search procedure and tuning parameters as GLMBoost-Cox.

**CoxBoost**

The likelihood-based Cox boosting model was implemented with the *CoxBoost* R package [Binder2008]. Internal cross-validation was used to determine the optimal number of boosting steps. We used `penalty = "optimCoxBoostPenalty"` with

`maxstepno = 500`, `K = 5`, `standardize = TRUE`, and `return.score = FALSE`. The internal routine *CoxBoost::optimCoxBoostPenalty()* identifies the penalty leading to the optimal number of boosting steps, which is then used to fit the final model.

**CoxLasso**

We implemented a penalized Cox model using *glmnet::cv.glmnet()* [Simon2011] with Lasso regularization. The model was fit with `family = "cox"`, `alpha = 1`, `standardize = TRUE`, `nfolds = 5`, `type.measure = "C"`, and `s = "lambda.min"`. Internal cross-validation determines the optimal penalization parameter `lambda`. Data subsampling–model pairs for which the selected feature set $|S_{ij}|$ was empty were removed. This occurred when CoxLasso produced unstable or degenerate fits, in which penalization eliminated all features. We did not use the `lambda.1se` option, as this would have exacerbated empty selections by prioritizing even sparser models at the cost of performance.

Feature selection methods

We considered the following feature selection methods:

- **CoxLasso (plain)**: N=1, B=0 — baseline embedded selection without resampling.
- **hEFS (9 models)**: N=9, B=100 — full ensemble using all nine survival learners.
- **hEFS (3 RSFs)**: N=3, B=100 — reduced ensemble with the two RSF variants and AORSF.
- **EFS (CoxLasso)**: N=1, B=100 — CoxLasso with embedded selection applied across 100 subsamples.
- **No feature selection (no FS)**: used only for the analysis in **Sup. Fig. 9**.

Integration Models

Choice of Model $M$ in **Fig. 2**. We compared three integration strategies: **BlockForest** (group-aware, **Fig. 5b & Sup. Fig. 9**), and two group-naïve methods (**RSF** and **CoxLasso, Sup. Fig. 8**). All models were trained on the concatenated clinical + multi-omics feature matrix after late fusion. Multi-omics data were standardized, and clinical variables were included in the joint matrix without prior feature selection and no penalization during the integration.

**BlockForest**

We implemented a group-aware Random Survival Forest using *BlockForest::blockfor()* [Hornung2019]. Models were trained with `splitrule = "logrank"`, `num.trees = 2000`, `num.trees.pre = 1000`, `nsets = 300` (default value of randomly generated hyperparameter modality weight candidates), `always.select.block = 0`, and `block.method = "BlockForest"`. Block structures corresponded to predefined omic groups. Note that BlockForest is equivalent to a standard RSF when applied to a single modality (e.g., GEX data in **Sup. Fig. 9**).

**RSF**

We implemented a Random Survival Forest using the *ranger* R package [Wright2017]. Models were trained with `splitrule = "logrank"`, `num.trees = 2000` (increased to stabilize performance estimates), and `importance = "none"` (since no RFE was used).

**CoxLasso**

We implemented a penalized Cox model using *glmnet::cv.glmnet()* [Simon2011] with Lasso regularization. The model was fit with `family = "cox"`, `alpha = 1`, `standardize = FALSE` (data already standardized), `nfolds = 5`, `type.measure = "deviance"`, `grouped = TRUE` (following the setup of [Wissel2023]), and `s = "lambda.min"` (which performs internal cross-validation to determine the optimal penalization parameter $\lambda$).

Evaluation

**Sparsity**

For each omic, sparsity was quantified by the **number of selected features**, $|S_{hEFS}|$ for any hEFS variant or $|S_{CoxLasso}|$ for plain CoxLasso (**Fig. 3a**). For the combined multi-omics panel, sparsity was defined as the total number of selected features across all modalities (including clinical variables), i.e. $|S_{ALL}| = \sum_{mod} |S_{mod}|$ (**Fig. 3b**).

**Stability**

Stability was evaluated using the **Nogueira similarity score** [Nogueira2018] (**Fig. 4a**, **Sup. Fig. 5**). Let $Z = \{S_1, S_2, ..., S_m\}$ denote the collection of feature sets obtained from

$m = 100$ Monte-Carlo resamplings for a given omics dataset after applying any FS method. Each set satisfies $S_i \subset \{1,...,p\}$ with $0 < |S_i| < p$. The score is defined as:

$$\varphi_{nog}(Z) = 1 - \frac{\frac{1}{p}\sum_{j=1}^{p}\frac{m}{m-1}p_j(1-p_j)}{\frac{\bar{k}}{p}\left(1-\frac{\bar{k}}{p}\right)}$$

where $p_j = \frac{1}{m}\sum_{i=1}^{m} 1\{j \in S_i\}$ is the observed selection frequency of feature $j$, and $\bar{k} = \frac{1}{m}\sum_{i=1}^{m}|S_i|$ is the average number of selected features across the $m$ resamplings. The score takes values in $[0, 1]$, with higher values indicating greater stability across resamplings. We used the implementation provided in the *stabm* R package [Bommert2021].

**Redundancy**

Let $S = \{f_1, f_2,..., f_N\}$ be the non-empty set of selected features ($N = |S|$) after applying any FS method on an omic dataset. The **redundancy rate (RR)** is defined as the mean absolute correlation across the unique feature pairs:

$$RR = \frac{2}{N(N-1)} \sum_{1 \leq i < j \leq N} |r_{ij}|$$

where $r_{ij}$ is the correlation between two features $f_i, f_j \in S$. Correlation can be measured using Pearson, Spearman, or the $\xi$ correlation coefficient [Chatterjee2021] (**Sup. Fig. 7**). For $\xi$-correlation, which is asymmetric ($\xi(x, y) \neq \xi(y, x)$), we followed the authors' recommendation to take the maximum value for each pair, i.e. $\xi_{ij} = max(\xi(f_i, f_j), \xi(f_j, f_i))$.

The **significant redundancy proportion (SRP)** was defined as the fraction of unique feature pairs $(f_i, f_j)$ exhibiting a statistically significant correlation (e.g., FDR-adjusted $p_{ij} < 0.05$; **Fig. 4b**, **Sup. Fig. 6**):

$$SRP = \frac{\#\{(f_i,f_j) : p_{ij} < 0.05\}}{\frac{N(N-1)}{2}}$$

Together, RR captures the average strength of pairwise correlations, while SRP quantifies the prevalence of statistically significant redundancies.

**Predictivity**

All models, both within hEFS and during multi-omic integration, predict a continuous risk score—either the linear predictor from Cox or AFT models, or the sum of the cumulative hazards (expected mortality) from RSF-type models [Ishwaran2008, Sonabend2022]. To assess the discriminatory ability of these predictions, we use **Harrell's concordance index (C-index)** [Harrell1982], defined as the probability that, for a randomly chosen pair of comparable patients $(i, j)$, the model assigns a higher risk score to the patient with the shorter survival time:

$$C = Pr(r_i > r_j \mid T_i < T_j)$$

Here, a pair is $(i, j)$ *comparable* if the patient with shorter observed time experienced the event (i.e., was not censored before $T_i$). The C-index therefore represents the proportion of correctly ordered comparable pairs and is the standard metric for evaluating survival discrimination. We used the implementation provided in the *mlr3proba* R package [Sonabend2021].

## Software and Data Availability

**Platform.** All benchmark experiments were performed on the Fox High Performance Computing (HPC) cluster for Educloud Research users, hosted by the University of Oslo IT Department [Fox2025]. Four interactive nodes were available, each with 2×AMD EPYC 7702 processors (64 cores/CPU, 2.0 GHz base, up to 3.35 GHz boost) and 1 TB RAM. All benchmarks were executed using 40 cores in parallel.

**Reproducibility and Availability**. All the codes, preprocessed datasets with metadata, and scripts to reproduce the presented results are openly available under the MIT license at https://github.com/bblodfon/pdac-efs-bench2024. We have further employed the *renv* R package [Ushey2025] to snapshot and restore package dependencies, ensuring full reproducibility.

**Software.** All analyses were implemented in R. Components of the *mlr3* R ecosystem were used for feature selection, learning algorithms, callbacks, and benchmarking, with additional functionality developed or extended specifically for this study. The following packages and survival models were used:

- **Core packages:**
  - *mlr3* (0.23)
  - *mlr3proba* (0.7.1) [Sonabend2021]
  - *mlr3extralearners* (1.0.0) [Fischer2025]

- *mlr3tuning* (1.2.1) [Becker2024]
- *mlr3fselect* (1.3.0) [Becker2025]
- *mlr3pipelines* (0.7.1) [Binder2021]
- **Survival models:**
  - *ranger* (0.17.0) [Wright2017]
  - *aorsf* (0.1.5) [Jaeger2022]
  - *xgboost* (1.7.8.1) [Chen2016, Barnwal2022]
  - *glmnet* (4.1-8) [Simon2011]
  - *mboost* (2.9-11) [Schmid2008, Hofner2014]
  - *CoxBoost* (1.5) [Binder2008]
  - *BlockForest* (0.2.6) [Hornung2019]

# Supplementary Information

## Supplementary Figure 1

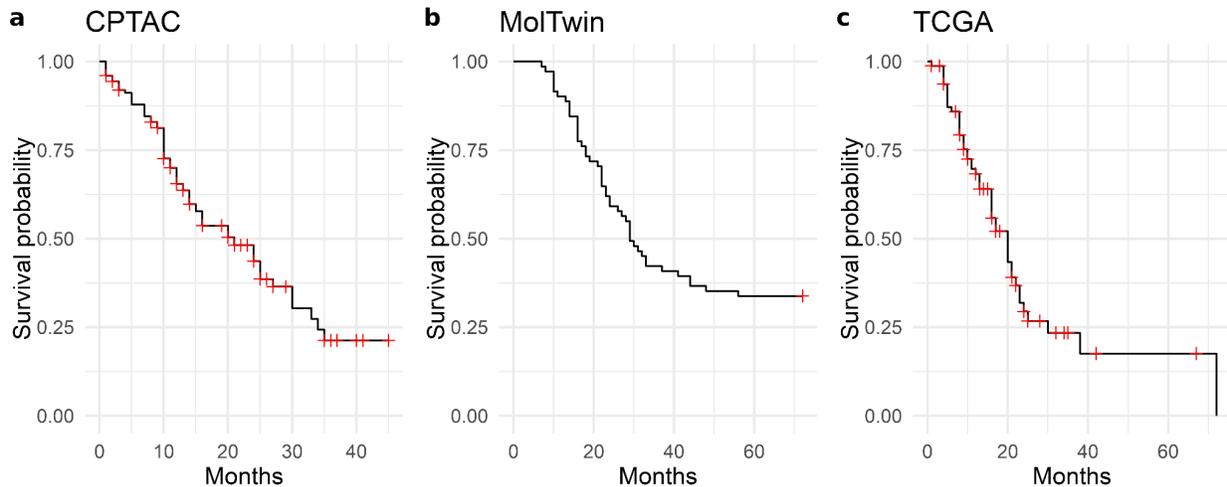

**Supplementary Figure 1. Kaplan–Meier survival curves for the three PDAC cohorts used in this study.** (a) CPTAC-PDAC [Cao2021] cohort  (b) MolTwin [Osipov2024] cohort; (c) TCGA-PDAC [Wissel2023] cohort. Each plot displays the estimated survival probability over time (in months) for patients with pancreatic ductal adenocarcinoma (PDAC). Red ticks indicate censored observations. The MolTwin cohort exhibits administrative censoring at the study's maximum follow-up time of 72 months, whereas censoring in the TCGA and CPTAC cohorts is distributed throughout the follow-up period.

# Supplementary Figure 2

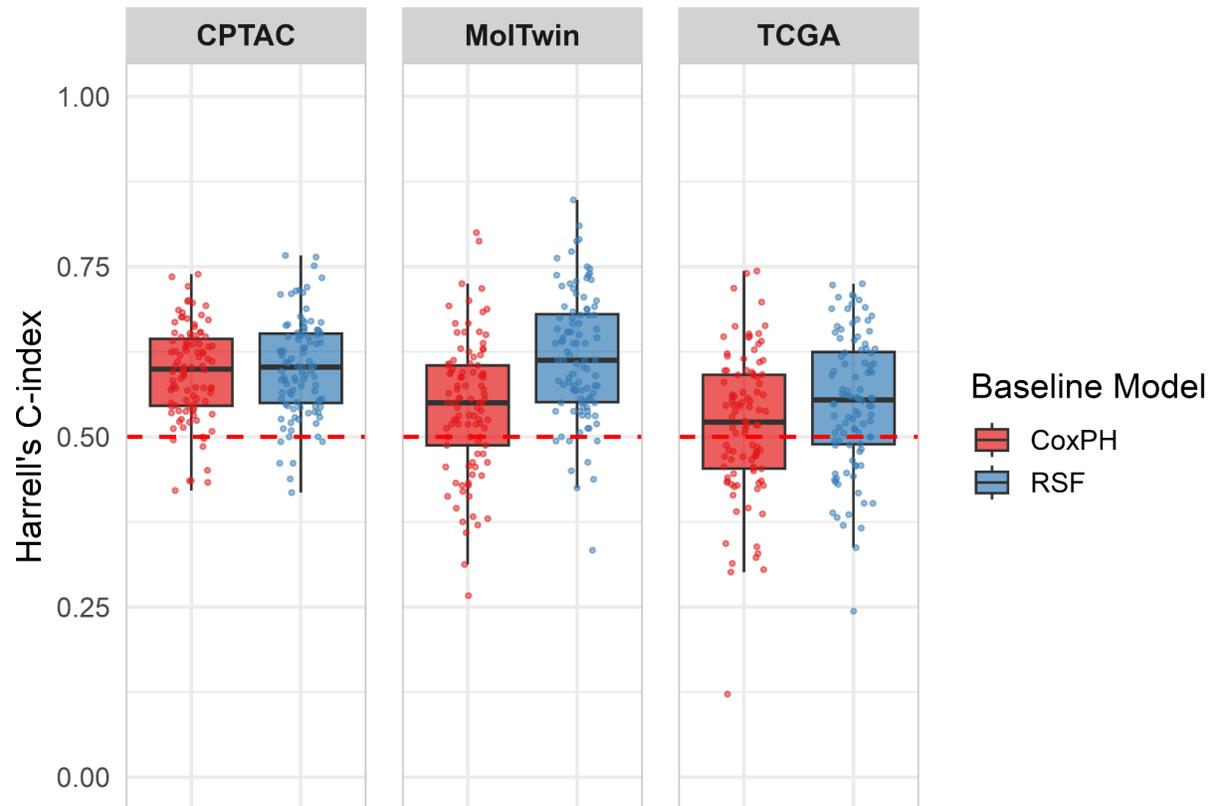

**Supplementary Figure 2. Comparison of baseline survival models trained on clinical features only.** Harrell's C-index distribution for Cox proportional hazards (CoxPH) and random survival forest (RSF) models trained exclusively on clinical variables, across 100 Monte Carlo cross-validation (MC-CV) iterations. Results are shown separately for the CPTAC [Cao2021], MolTwin [Osipov2024] and TCGA [Wissel2023] PDAC cohorts. RSF consistently outperforms CoxPH in two out of three datasets, motivating its use as the reference integration model in our benchmark. The red dashed line marks the random discriminatory performance with a C-index of 0.5.

# Supplementary Figure 3

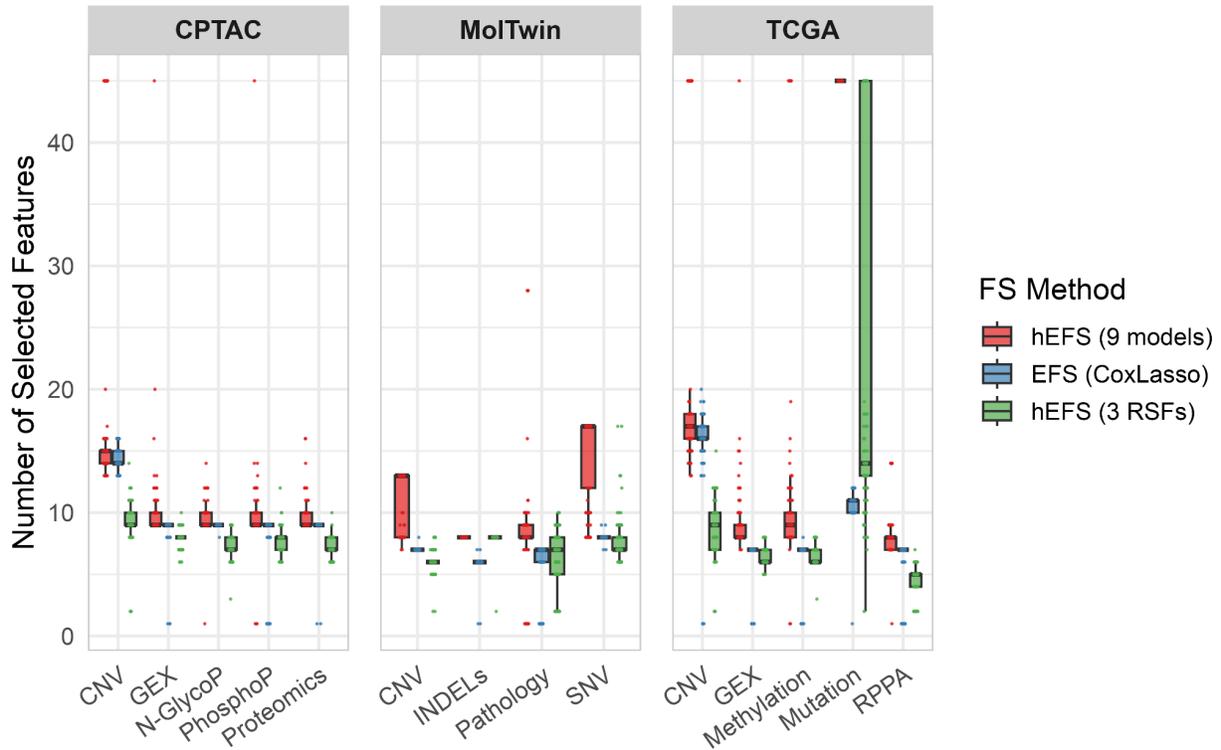

**Supplementary Figure 3. Per-omic feature sparsity across three PDAC multi-omics datasets, excluding the CoxLasso baseline.** This figure replicates the analysis from **Fig. 3a** but excludes the CoxLasso baseline to better visualize the lower feature counts of the hybrid ensemble feature selection (hEFS) methods. Across 100 Monte Carlo cross-validation iterations, the hEFS (3 RSF) yields the most sparse selections, followed by the EFS (CoxLasso), and then the full hEFS with all nine models—regardless of the original omic dimensionality. In most cases, fewer than 15 features on average are selected per omic layer.

# Supplementary Figure 4

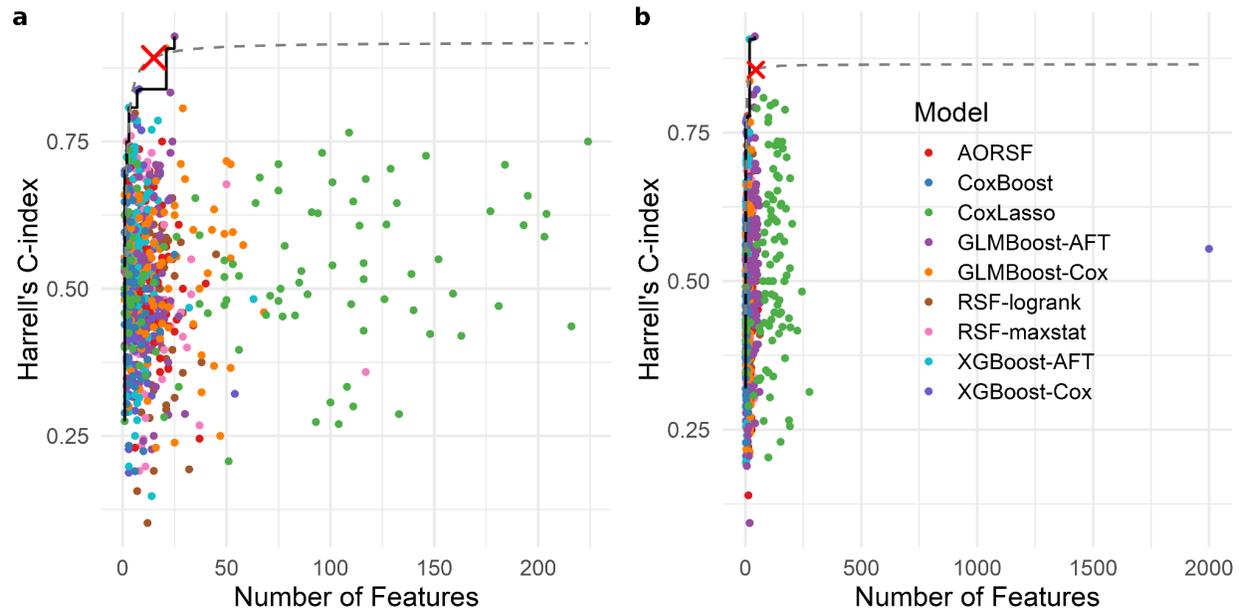

**Supplementary Figure 4. Biomarker set size variability driven by the shape of the estimated Pareto front.** Pareto fronts (empirical: stepwise, black; estimated via inverse-number-of-features weighting: dashed, grey; **Methods**) and selected knee points (red crosses) from two outer resampling iterations using mutation data from the TCGA cohort [Wissel2023]. Each case corresponds to 65 patients and the 2000 most variable features. These illustrate how the number of selected features in the hEFS (9 models) variant can vary depending on the structure of the Pareto front. **(a)** The Pareto front yields a compact solution with a knee point at 15 features. Points not on the empirical front—such as those from CoxLasso—are excluded from the estimated front, so the knee point is calculated solely using the remaining Pareto-optimal solutions (grey dashed line). **(b)** The inclusion of a model–data pair with substantially more features (e.g., XGBoost-Cox in one inner resampling) stretches the estimated Pareto front, shifting the knee point to a higher value (45 features). This highlights how such outliers can influence the Pareto front geometry and increase the selected biomarker set size.

# Supplementary Figure 5

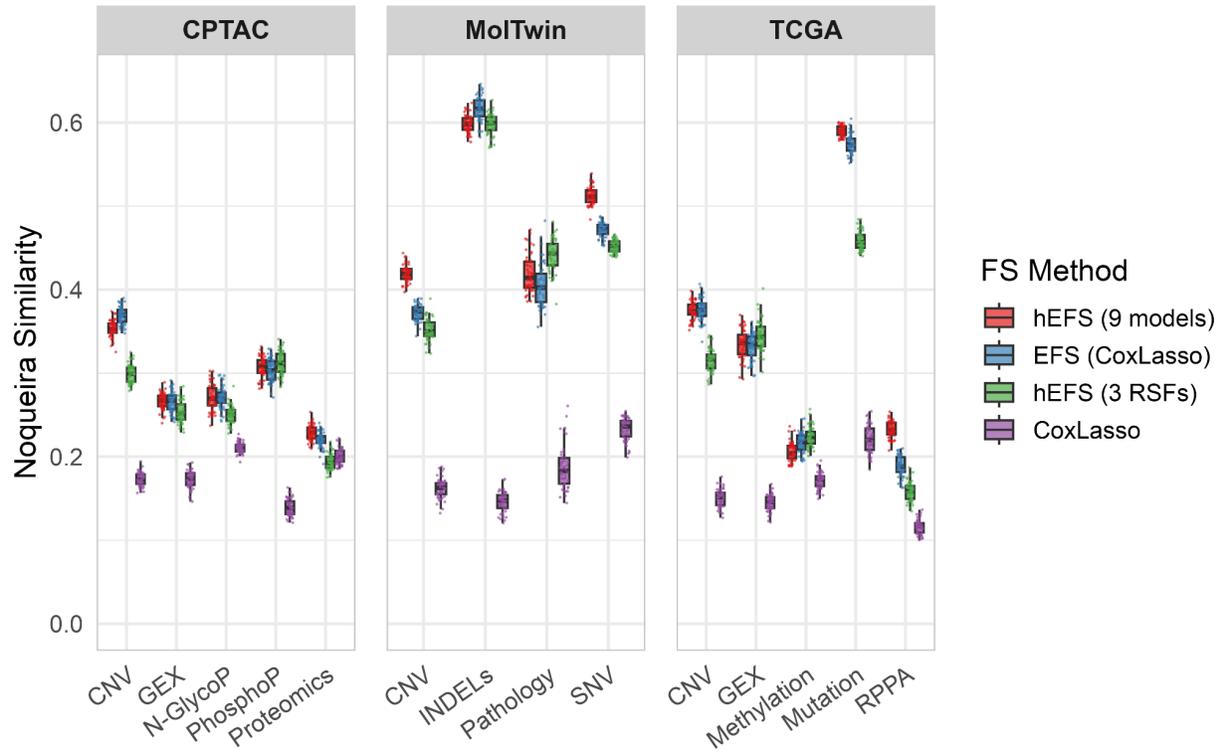

**Supplementary Figure 5. Robustness of stability estimates across Monte Carlo cross-validation (MC-CV) iterations.** For each dataset, omic type, and feature selection method combination, we randomly sampled 50 MC-CV runs (and their corresponding selected feature subsets) from the original 100 runs, and repeated this resampling procedure 50 times. Nogueira similarity scores were computed for each resampled set, and boxplots summarize the distribution of these scores across replicates. The consistently low variability (interquartile ranges typically <0.05) confirms the robustness of the stability patterns observed in **Fig. 4a**.

# Supplementary Figure 6

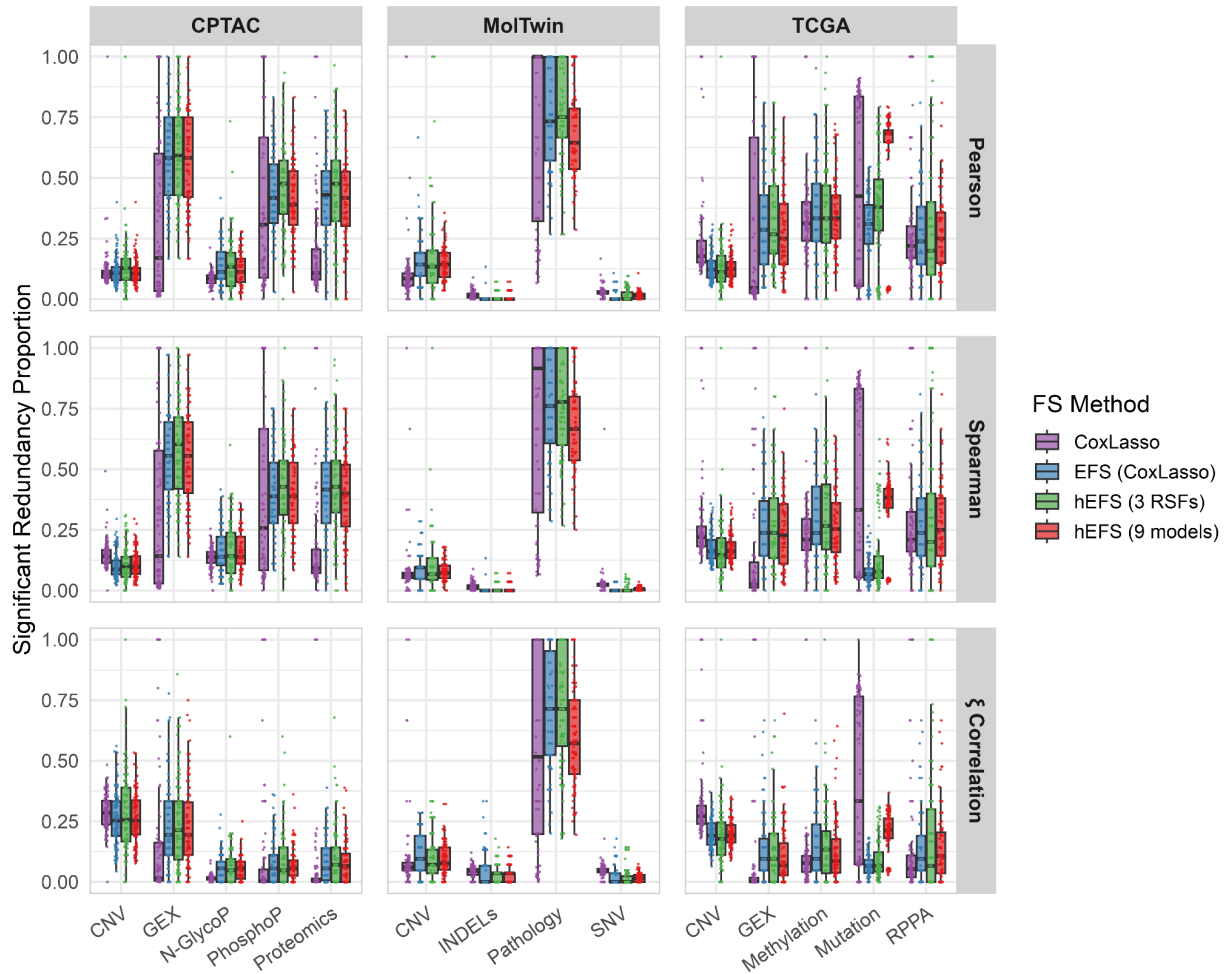

**Supplementary Figure 6. Significant Redundancy Proportion across Omic Types and Correlation Metrics.** Each panel shows the proportion of feature pairs identified as significantly redundant (FDR-adjusted $p < 0.05$) for each feature selection (FS) method, stratified by omic type (x-axis) and PDAC cohort (columns). Redundancy was assessed using three different correlation measures (rows): Pearson, Spearman, and ξ correlation [Chatterjee 2021]. Trends across correlation metrics were consistent, with FS methods showing broadly similar relative patterns within each omic type. Notably, ξ correlation (bottom row; same as in **Fig. 4b**) yielded consistently lower redundancy proportions across omics and datasets, reflecting its sensitivity to more general forms of statistical dependence rather than strict linear or monotonic associations. Boxplots summarize variability across 100 Monte Carlo cross-validation iterations.

# Supplementary Figure 7

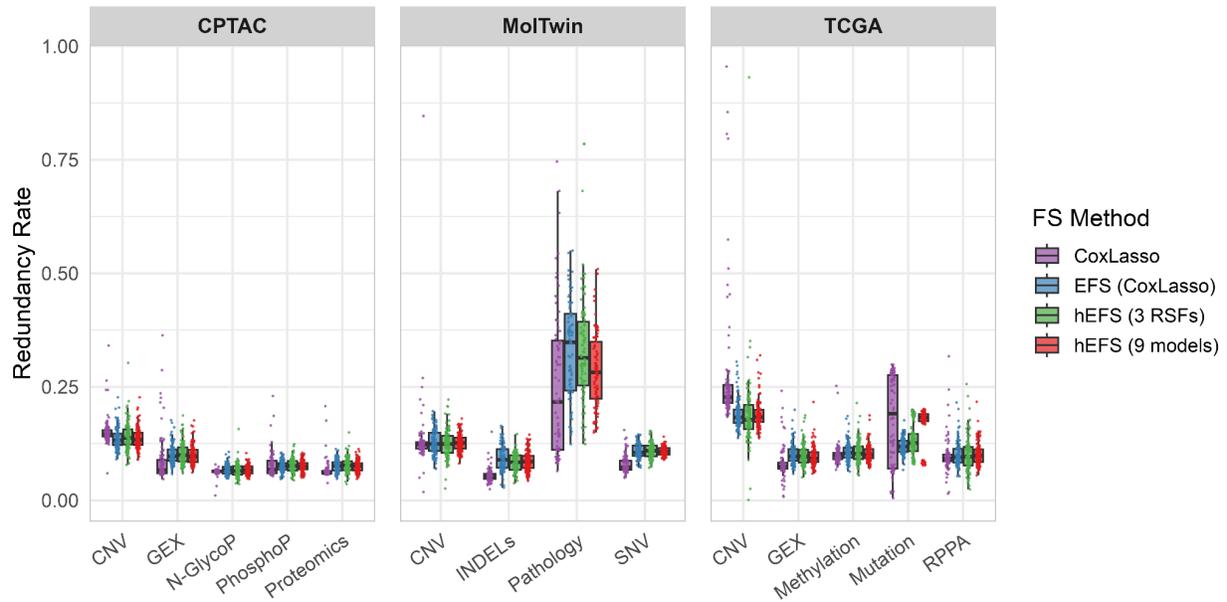

**Supplementary Figure 7. Redundancy rate across omic types based on mean absolute ξ correlation.** Boxplots show the distribution of mean absolute ξ correlation scores across 100 Monte Carlo cross-validation iterations, stratified by feature selection (FS) method, omic type, and PDAC cohort. Lower values indicate lower average redundancy among selected feature pairs. Across most omics and cohorts, redundancy rates remained low (median ≤ 0.2) and comparable between FS methods.

# Supplementary Figure 8

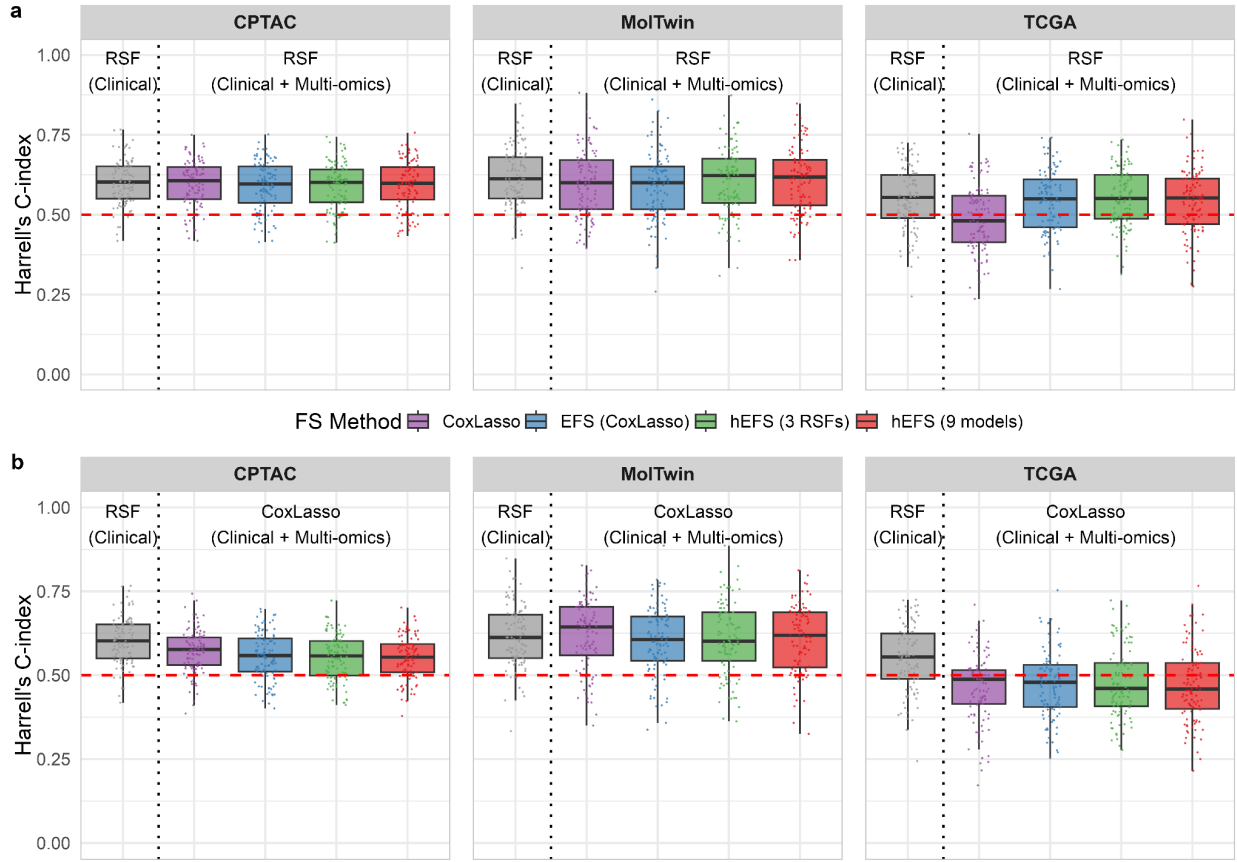

**Supplementary Figure 8. Predictive performance across feature selection methods and PDAC cohorts using group-naive integration models.** Discriminatory performance of (a) Random Survival Forests (RSF) and (b) CoxLasso models, that integrate clinical data with late-fused, feature-selected multi-omics matrices, benchmarked against a baseline RSF trained only on clinical features (grey). The red dashed line indicates random discriminatory ability (C-index = 0.5). Higher C-index, better discriminatory performance. Boxplots summarize variability across 100 Monte Carlo cross-validation iterations.

# Supplementary Figure 9

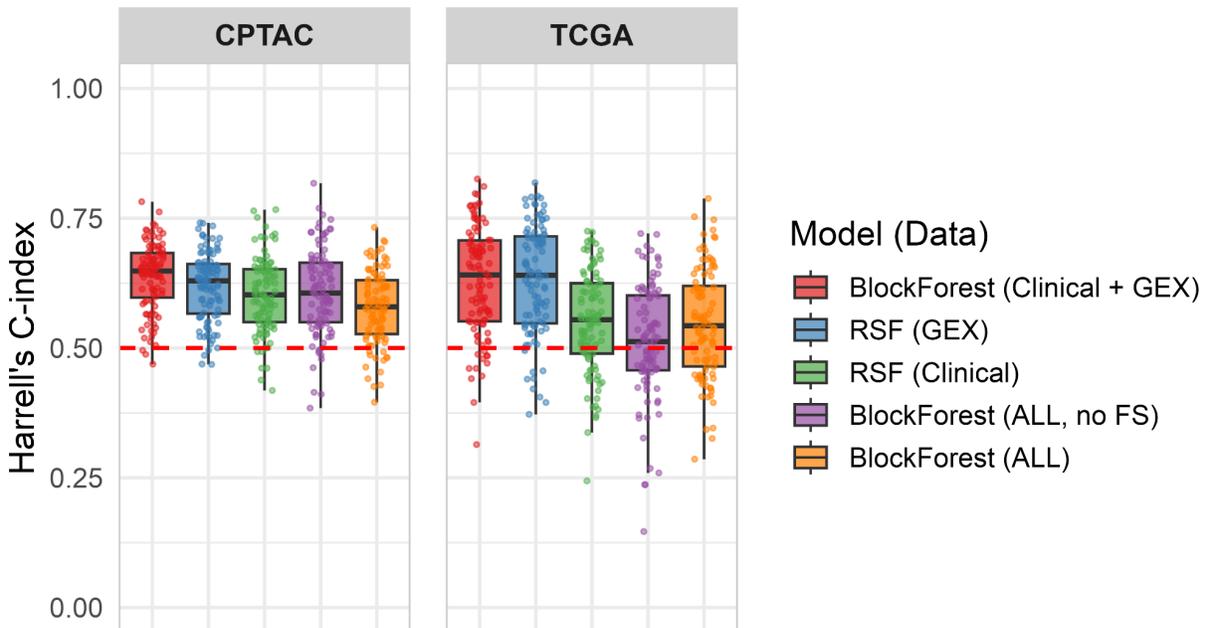

**Supplementary Figure 9**. Comparison of discriminatory performance between single-omic (gene expression), clinical-only, clinical + GEX, and multi-omic models across the CPTAC [Cao2021] and TCGA [Wissel2023] PDAC cohorts. 'ALL' here refers to clinical + all available multi-omics data for a particular cohort. Feature selection with hEFS (all 9 models) was applied to the GEX data and used in RSF (GEX; blue), BlockForest (Clinical + GEX; red), and BlockForest (ALL; orange). For comparison, BlockForest (ALL, no FS) used the full set of omics plus clinical data without feature selection. Boxplots summarize variability across 100 Monte Carlo cross-validation iterations. The MolTwin cohort is not shown, as GEX data were not used due to sample size constraints.

# Supplementary Table 1

**Supplementary Table 1.** Average execution times (seconds) with standard deviations for feature selection methods across PDAC cohorts (ordered by decreasing sample size), averaged over omic data types and Monte Carlo CV iterations.

| PDAC Dataset | CoxLasso | EFS (CoxLasso) | hEFS (3 RSFs) | hEFS (9 models) |
|---|---|---|---|---|
| CPTAC [Cao2021] | 2.40 ± 1.50 | 16.99 ± 2.62 | 115 ± 32 | 836 ± 182 |
| TCGA [Wissel2023] | 2.24 ± 1.02 | 13.79 ± 1.98 | 77 ± 16 | 634 ± 99 |
| MolTwin [Osipov2024] | 1.72 ± 0.67 | 11.76 ± 3.00 | 50 ± 16 | 473 ± 86 |